\documentclass[prd,superscriptaddress,showpacs,onecolumn]{revtex4}
\usepackage{amssymb}
\usepackage{bm}
\usepackage{amsfonts}
\usepackage{latexsym}
\usepackage[latin1]{inputenc}
\usepackage{graphicx}
\usepackage{amsmath}
\usepackage{palatino}
\usepackage{mathpazo}
\usepackage{textcomp}
\usepackage{float}
\usepackage{booktabs}
\usepackage{dcolumn}
\usepackage{hyperref}
\usepackage{amsmath}
\usepackage{xcolor}
\usepackage{mathtools}
\usepackage{orcidlink}
\usepackage[caption=false]{subfig}
\usepackage{commath}

\hypersetup{colorlinks,citecolor=blue}

\def\jnl@style{\it}
\def\aaref@jnl#1{{\jnl@style#1}}
\def\aaref@jnl#1{{\jnl@style#1}}
\def\aj{\aaref@jnl{AJ}}
\def\apj{\aaref@jnl{ApJ}}
\def\apjl{\aaref@jnl{ApJ}}
\def\apjs{\aaref@jnl{ApJS}}
\def\apss{\aaref@jnl{Ap\&SS}}
\def\aap{\aaref@jnl{A\&A}}
\def\aapr{\aaref@jnl{A\&A~Rev.}}
\def\aaps{\aaref@jnl{A\&AS}}
\def\mnras{\aaref@jnl{Mon.~Not.~Roy.~Astron.~Soc.}}
\def\prd{\aaref@jnl{Phys.~Rev.~D}}
\def\prc{\aaref@jnl{Phys.~Rev.~C}}
\def\prl{\aaref@jnl{Phys.~Rev.~Lett.}}
\def\qjras{\aaref@jnl{QJRAS}}
\def\skytel{\aaref@jnl{S\&T}}
\def\ssr{\aaref@jnl{Space~Sci.~Rev.}}
\def\zap{\aaref@jnl{ZAp}}
\def\nat{\aaref@jnl{Nature}}
\def\aplett{\aaref@jnl{Astrophys.~Lett.}}
\def\apspr{\aaref@jnl{Astrophys.~Space~Phys.~Res.}}
\def\physrep{\aaref@jnl{Phys.~Rep.}}
\def\physscr{\aaref@jnl{Phys.~Scr}}
\def\commat{\aaref@jnl{Comm.~Math.~Phys.}}
\def\science{\aaref@jnl{Science}}
\def\cqg{\aaref@jnl{Classical Quant.~Grav.}}
\def\jpcs{\aaref@jnl{JPCS}}
\def\ijmpd{\aaref@jnl{Int.~J.~Mod.~Phys.~D}}
\def\grg{\aaref@jnl{Gen.~Relat.~Gravit.}}
\def\rpp{\aaref@jnl{Rep.~Prog.~Phys.}}
\def\npa{\aaref@jnl{Nucl.~Phys.~A}}
\def\lrr{\aaref@jnl{Living Rev.~Rel.}}
\def\jcap{\aaref@jnl{J.~Cosmology Astropart.~Phys.}}
\def\rmp{\aaref@jnl{Rev.~Mod.~Phys.}}

\allowdisplaybreaks[1]

\begin{document}

\noindent Phys. Dark Universe {\bf 32}, 100804 (2021) \hfill  \url{https://doi.org/10.1016/j.dark.2021.100804}

\title{Late-time acceleration with a scalar field source: Observational
constraints and statefinder diagnostics}
\author{S. K. J. Pacif\orcidlink{0000-0003-0951-414X}}
\email{shibesh.math@gmail.com}
\affiliation{Department of Mathematics, School of Advanced Sciences, Vellore Institute of
Technology, Vellore 632014, Tamil Nadu, India.}
\author{Simran Arora\orcidlink{0000-0003-0326-8945}}
\email{dawrasimran27@gmail.com}
\affiliation{Department of Mathematics, Birla Institute of Technology and Science-Pilani, 
\\
Hyderabad Campus, Hyderabad-500078, India.}
\author{P.K. Sahoo\orcidlink{0000-0003-2130-8832}}
\email{pksahoo@hyderabad.bits-pilani.ac.in}
\affiliation{Department of Mathematics, Birla Institute of Technology and Science-Pilani, 
\\
Hyderabad Campus, Hyderabad-500078, India.}

\begin{abstract}
This article discusses a dark energy cosmological model in the standard
theory of gravity - general relativity with a broad scalar field as a
source. Exact solutions of Einstein's field equations are derived by
considering a particular form of deceleration parameter $q$, which shows a
smooth transition from decelerated to accelerated phase in the evolution of
the universe. The external datasets such as Hubble ($H(z)$) datasets,
Supernovae (SN) datasets, and Baryonic Acoustic Oscillation (BAO) datasets
are used for constraining the model par parameters appearing in the
functional form of $q$. The transition redshift is obtained at $%
z_{t}=0.67_{-0.36}^{+0.26}$ for the combined data set ($H(z)+SN+BAO$), where
the model shows signature-flipping and is consistent with recent
observations. Moreover, the present value of the deceleration parameter
comes out to be $q_{0}=-0.50_{-0.11}^{+0.12}$ and the jerk parameter $%
j_{0}=-0.98_{-0.02}^{+0.06}$ (close to 1) for the combined datasets, which
is compatible as per Planck2018 results. The analysis also constrains the
omega value i.e., $\Omega _{m_{0}}\leq 0.269$ for the smooth evolution of
the scalar field EoS parameter. It is seen that energy density is higher for
the effective energy density of the matter field than energy density in the
presence of a scalar field. The evolution of the physical and geometrical
parameters is discussed in some details with the model parameters' numerical
constrained values. Moreover, we have performed the state-finder analysis to
investigate the nature of dark energy.
\end{abstract}

\maketitle

%\color{red}
\color{black} %% For one column

%%%%%%%%%%%%%%%%%%%%%%%%%%%%%%%%%%%%  DATE  %%%%%%%%%%%%%%%%%%%%%%%%%%%%%%%%%%%%
%\date{\today}

%\input epsf.tex
%%%%%%%%%%%%
%%%%%%%%%%%

\section{Introduction}\label{I}

Cosmological observations indicate that our universe is going through a
phase of an accelerated expansion \cite{Riess,Perlmutter}, which is also
supported by the recent SNe Ia observations \cite{SNIa}, CMB observations 
\cite{CMB}, BAO peak experiments \cite{BAO} and $H(z)$ measurements \cite{Hz}%
. These observations also indicate that the cosmological entity is
responsible for the acceleration. In addition to that, this entity should
also create an anti-gravitational effect to push the universe apart. The
unknown force responsible for the accelerated expansion possessing a
negative pressure is generally termed "dark energy"(DE). According to the $%
\Lambda$CDM model, the best current measurement for dark energy is 69
percentage, i.e., $\frac{3}{4}$th of the total energy in the present-day
observable universe. Since ordinary baryonic matter does not have such an
equation of state, let alone to account for such a lion-share of the energy
budget of the universe, several alternate scenarios have been proposed and
investigated\cite{alternate}.\newline

Since dark energy is mysterious and not much idea about its nature, several
dark energy candidates have been proposed. Out of these various possible
choices to study dark energy, Einstein's cosmological constant ($\Lambda $)
introduced in 1917, serves the best and simplest candidate, as described in
the literature. This suggests that the repulsive nature of $\Lambda $ is
responsible for accelerating the universe with the equation of state $\omega
=-1$. However, this authentic candidate suffers from some long-standing
cosmological constant problem and also the constant equation of state. As we
know, the EoS parameter is the connection between energy density and
pressure i.e. $\omega = \frac{p}{\rho}$. The EoS parameter is used to
characterize the universe's decelerated and accelerated expansion. It
categorizes different phases of the universe as, If $\omega=\frac{1}{3}$,
the model indicates the phase dominated by radiation, while $\omega=0$
represents the phase dominated by matter. In the current accelerated period
of evolution, the quintessence period is shown by $-1<\omega\leq0$ and the
cosmological constant $\omega=-1$, i.e., the $\Lambda$CDM model and the
phantom age by $\omega<-1$. The recent fine-tuning problem can be minimized
by considering the equation of state as time-dependent. One such model
having this property is the scalar field, also known as quintessence \cite%
{Sahni}. The presence of scalar fields $\phi$ is predicted by several
fundamental physics theories, encouraging to study the dynamic properties of
scalar fields in cosmology. A wide range of scalar-field dark energy models
has been suggested so far. Among several, these include quintessence,
K-essence, tachyon, phantoms etc. Quintessence depends on scalar fields'
potential energy to contribute to the acceleration of the late time
universe. The potential $V(\phi )$ has the property such that scalar fields
are rotating down potential approach a common evolutionary path. For large
values of $\phi $, the potential becomes flat, winding scalar fields to slow
down, allowing the universe to accelerate. This ordinary scalar field model
is a viable alternative that has a dynamical equation of state (EoS) wherein
the EoS parameter $\omega $ ranges in between $-1/3$ to $-1$. A route to
dark energy is also provided by Chaplygin gas \cite{AY/2001} which has a
peculiar EoS and also phantom field \cite{Bamba/2012} for which $\omega $
crosses $-1$. \newline

Implementing SN distance measurements \cite{Riess, Perlmutter} with baryon
acoustic peak measurements in the power spectrum of cosmic microwave
background(CMB) suggests that our universe is accelerating and composed
predominantly of baryon, dark matter, and dark energy. The cosmic microwave
background (CMB) is landmark proof of the universe's Big Bang origin.
Precise CMB measurements are essential to cosmology, as any proposed model
of the universe must clarify this radiation. Current observational data is
used to constrain the $f(z)$ models and describe distance-redshift
relations. Here, $f(z)$ refers to constraining the $H(z)$ models.
Nevertheless, given its role as a hypothesis, the $\Lambda$CDM model was
immensely successful in explaining most cosmological observations. In
addition to a moderately significant difference with broad angular scale CMB
observations, $\Lambda$CDM provided an almost ideal match to measurements
made by the Wilkinson Microwave Anisotropy probe (WMAP) satellite project
mission. Even in parallel with corresponding observational data such as BAO
surveys, Type Ia supernovae, and Hubble constant direct measurements. The
Baryon Acoustic Oscillations (BAO) matter clustering provides a `standard
ruler' for length scale in cosmology the same way as supernovae offer a
`standard candle' for astronomical observations. There are different
observational data also discussed in \cite{Valentino,J.K.Singh}: Cosmic
microwave background radiation (CMB) acts as authentication of big bang
theory, Sloan digital sky survey (SDSS), which provides a map of the
distribution of the galaxy and encodes the existing variations in the
universe, Baryon acoustic oscillations (BAO) \cite{Aghanim} estimates
large-scale structures in the universe that make the dark energy more
attractive, QUASARS brings out the matter between observers and quasars, SNe
Ia observations are the instruments for measuring the cosmic distances known
as standard candles. Compilations of Hubble measurements\cite{Casertano} are
regarded as cosmic chronometers with a sample covering the redshift range of 
$0<z<1.97$. And the latest type 1048 SNe Ia covers the redshift range of $%
0.01<z<2.26$. Also, the luminosity distance data of 1048 type Ia supernovae
from Pantheon \cite{Scolnic} is recently developed.\newline

The current practice for finding the cosmic evolution is to develop the
model from observational data. The method of finding a viable cosmological
model is called reconstruction. Starobinsky \cite{Staro/1998}, in his
pioneering work, considered the scalar field potential, which was used as
the dark energy, and has reconstructed the cosmological model using the
density perturbation data. Later, the observational data of distance
measurement from supernova has been utilized in \cite{Huterer/1999,
Saini/2000}. The effective equation of dark energy through the
parametrization of the quintessence scalar field and potential is discussed
in \cite{Guo/2005}. In general, two types of reconstruction available in the
literature. The first one is based on the parametric form of the DE equation
of state $\omega _{DE}$ \cite{Mukherjee/2016} and the estimation of
parameters from the observational data. The second one is a non-parametric
formulation attempting to estimate the evolution of $\omega_{DE}$ directly
from the observational data without considering any parametric form \cite%
{nair/2014}.\newline

In cosmology, authors try to understand the cosmic acceleration through
analyzing the kinematic variables like the Hubble parameter ($H$), the
deceleration parameter ($q$), and the jerk parameter ($j$), where all these
parameters are derived from the derivatives of the scale factor $a$ \cite%
{Mamon/2018}. The kinematic approach is more advantageous as it does not
depend on any model-specific assumptions. It is described by some metric
theory of gravity and is assumed that the present Universe is isotropic and
homogeneous at cosmological scales \cite{parameter}. In the literature, one
can find many attempts to constrain the current values of $H, q$, and the
jerk $j$ by parametrizing the deceleration parameter $q$ \cite{dece}.\newline

In this work, we have considered the scalar field model in flat FLRW
space-time. We come across the two field equations along with the
conservation equations in the scalar field and matter field \cite{Sudipta,
Narayan}. Further, the proposed deceleration parameter is considered so that
the difficulty in solving these equations with four unknowns $a(t)$, $%
\rho_{\phi }$, $p_{\phi }$ and $V(\phi )$ can be reduced. Therefore, we
obtained the expressions for $\rho _{\phi }$, $p_{\phi }$. The model
parameters are constrained using the Hubble, BAO, and SN datasets. The
behavior of the equation of state parameter $\omega =\frac{p_{\phi }}{\rho
_{\phi }}$ and the deceleration parameter has been observed showing the
transition from decelerated to accelerated phase. The various kinematic
variables such as jerk, snap, and lerk are studied, indicating the
accelerated expansion. Further, the temporal evolution of dark energy
mimicked by our model has been shown by the statefinder diagnostics {r-s}
and {r-q} pairs. The two planes describe quintessence dark energy, the
Chaplygin gas model, and the $\Lambda $CDM model.\newline

The manuscript is organized as follows: In Section \ref{II}, we present the
Einstein field equations along with the scalar field. In Section \ref{III},
we derive the kinematic quantities from the second-degree parametrization of
the deceleration parameter and found the observational constraints on the
model parameters involved. The kinematic parameters of the cosmological
model are discussed in Section \ref{IV}. In Section \ref{V}, we present some
geometrical diagnostics. Finally, in Section \ref{VI}, we present our
results and conclusions.

\section{EFEs \& scalar field formalism}\label{II}

To begin our analysis, we consider the homogeneous and isotropic flat
Friedmann-Lemaitre-Robertson-Walker (FLRW) space-time as,

\begin{equation}  \label{1}
ds^{2}=dt^{2}-a(t)^{2}\left[ dr^{2}+r^{2}d\Omega ^{2}\right] ,
\end{equation}%
where $a(t)$ is the scale factor of the universe and $d\Omega ^{2}=d\theta
^{2}+\sin ^{2}\theta d\varphi ^{2}$.

For a general scalar field together with the cold dark matter as sources,
the Einstein field equations (EFEs) are obtained as \cite{Sudipta, Narayan}, 
\begin{equation}
3\left( \frac{\dot{a}}{a}\right) ^{2}=\rho _{m}+\frac{1}{2}\dot{\phi}%
^{2}+V(\phi ),  \label{2}
\end{equation}%
and 
\begin{equation}
2\frac{\ddot{a}}{a}+\left( \frac{\dot{a}}{a}\right) ^{2}=-\frac{1}{2}\dot{%
\phi}^{2}+V(\phi ).  \label{3}
\end{equation}%
Here, $\rho _{m}$ is the matter density, $\phi $ is the scalar field and $%
V(\phi )$ is the scalar field potential. The overhead dot denote the
derivative of the quantity with respect to the cosmic time `$t$'. The units
have been chosen in such a way that $8\pi G=c=1$. The energy density and
pressure due to the field $\phi $ are, 
\begin{equation}
\rho _{\phi }=\frac{1}{2}\dot{\phi}^{2}+V(\phi )\text{,}  \label{4}
\end{equation}

\begin{equation}
p_{\phi }=\frac{1}{2}\dot{\phi}^{2}-V(\phi )\text{.}  \label{5}
\end{equation}

The equation of state for the scalar field is given by, $\omega _{\phi }=%
\frac{p_{\phi }}{\rho _{\phi }}$. We also consider minimal interaction
between cold dark matter and the dark energy for which the conservation of
energy and momentum yield the continuity equations for matter and scalar
fields separately as, 
\begin{equation}
\dot{\rho _{m}}+3H\rho _{m}=0\text{,}  \label{7}
\end{equation}

\begin{equation}
\dot{\rho _{\phi }}+3H(\rho _{\phi }+p_{\phi })=0\text{,}  \label{6}
\end{equation}%
where $H=$ $\frac{\dot{a}}{a}$ is the Hubble parameter. On solving Eqs. (\ref%
{7}), we come across the solution for the matter energy density $\rho _{m}$
as, 
\begin{equation}
\rho _{m}=\rho _{0}a^{-3}.  \label{rhom}
\end{equation}

With the understanding, $\rho _{eff}=\rho _{\phi }+\rho _{m}$ and $%
p_{eff}=p_{\phi }+p_{m}$, from Eqs. (\ref{2}), (\ref{3}) and (\ref{rhom}),
we may obtain the general expressions for the energy densities and pressure
and potential function of the scalar field as, 
\begin{equation}
\rho _{eff}=3H^{2}\text{, }\rho _{\phi }=3H^{2}-\rho _{m}\text{,}  \label{10}
\end{equation}

\begin{equation}
p_{eff}=p_{\phi }=(2q-1)H^{2}\text{, (since }p_{m}=0\text{),}  \label{11}
\end{equation}%
\begin{equation}
V(\phi )=(2-q)H^{2}-\frac{\rho _{m}}{2}\text{,}  \label{8}
\end{equation}%
where $q=-\frac{a\ddot{a}}{\dot{a}^{2}}$ is the deceleration parameter.

\section{Kinematic variables $\&$ Observational constraints}\label{III}

The kinematic variables play an important role in a cosmological model study
e.g. the deceleration parameter describes the behavior of the universe
whether it is ever decelerating, ever accelerating or has any transition
phase or multiple transition phases etc. Similarly, the equation of state
parameter describes the physical significance of the energy sources in the
evolution of the universe. Also, the above system of field equations need
one more equation to close the system for the complete determination of
other cosmological parameters and for their evolutionary behavior e.g.
pressure, energy densities and EoS parameter, potential function. This
supplementary equation can be assumed as a functional form of any
cosmological parameter. Parameterizations of Hubble parameter, deceleration
parameter, EoS parameter etc. provide the necessary constraint equation (see 
\cite{pacif2016}). In this paper, we employ a generalized varying
deceleration parameter of the second degree introduced in \cite{Bakry and
Shafeek} of the form, 
\begin{equation}
q(t)=(8\alpha ^{2}-1)-12\alpha t+3t^{2},  \label{13}
\end{equation}%
where $\alpha $ is an arbitrary constant and constrained from a $\chi ^{2}$
test using some external observational datasets. The corresponding Hubble
parameter reads, 
\begin{equation}
H(t)=\frac{1}{t(2\alpha -t)(4\alpha -t)}.  \label{14}
\end{equation}

Integrating the above equation, we get the explicit expression of scale
factor as, 
\begin{equation}  \label{15}
a(t)=\beta \frac{\lbrack t(4\alpha -t)]^{\frac{1}{8\alpha ^{2}}}}{(2\alpha
-t)^{\frac{1}{4\alpha ^{2}}}},
\end{equation}%
where $\beta $ is an integrating constant.\newline
\qquad As we are interested in studying the late-time universe, we should
express the above geometrical parameters in terms of redshift $z$ related to
scale factor as $a(t)=1/(1+z)$. So, the kinematic quantities $H(z)$ and $%
q(z) $ are expressed as functions of redshift as, 
\begin{equation}  \label{16}
H(z)=\frac{H_{0}\left( (\beta (z+1))^{8\alpha ^{2}}+1\right) ^{3/2}}{\left(
\beta ^{8\alpha ^{2}}+1\right) ^{3/2}(z+1)^{4\alpha ^{2}}},
\end{equation}%
\begin{equation}  \label{17}
q(z)=-\frac{(\beta (z+1))^{8\alpha ^{2}}+\alpha ^{2}\left( 4-8(\beta
(z+1))^{8\alpha ^{2}}\right) +1}{(\beta (z+1))^{8\alpha ^{2}}+1}.
\end{equation}

The expressions for $H(z)$ and $q(z)$ in Eqs. \eqref{16} and \eqref{17}
contain two free parameters $\alpha $ and $\beta $ (let us call them model
parameters). As we can see, the evolution depends on the values of model
parameters $\alpha $ and $\beta $, their values should be chosen properly to
describe the current evolution. So, we will constrained their values through
some observational datasets.

As we know, the modern cosmology is heavily dependent on observations and
describe the validation of any theoretical model obtained and also find
constraints on the model parameters, here in this study, we find
observational constraints on our model parameters $\alpha $ \& $\beta $
using observational Hubble datasets (OHD) containing a sample of $57$ data
points, Type Ia supernovae datasets (known as standard candles, used to
measure the expansion of Universe) containing a sample of $580$ data points
(Union$2.1$ compilation datasets) and Baryon Acoustic Oscillations (BAO)
datasets (used to measure the structure in the Universe).

\subsection{OHD sample}

\qquad A list of $57$ points of Hubble parameter data in the redshift range $%
0.07\leqslant z\leqslant 2.42$ is compiled by Sharov and Vasiliev \cite%
{sharov} (see the Appendix in \cite{sharov}) is considered here. We also
take a prior for the present value of the Hubble constant from Planck 2018
results \cite{Hz-Plank} as $H_{0}=67.8$ $Km/s/Mpc$ to complete the data set.
The mean values of the model parameters $\alpha $ \& $\beta $ are determined
by minimizing the chi square value (which is equivalent to the maximum
likelihood analysis). The chi square value is given by,

\begin{equation}  \label{chihz}
\chi _{OHD}^{2}(p_{s})=\sum\limits_{i=1}^{28}\frac{%
[H_{th}(p_{s},z_{i})-H_{obs}(z_{i})]^{2}}{\sigma _{H(z_{i})}^{2}},
\end{equation}
where, $H_{th}$ and $H_{obs}$ respectively refers to the theoretical and
observed value of Hubble parameter $H$ and $p_{s}$ refers to the parameters
of the model to be constrained. $\sigma _{H(z_{i})}$ stands for the standard
error in the observed value of $H$.

\subsection{Union 2.1 compilation datasets sample}

\qquad For our analysis, we have used the Union $2.1$ compilation supernovae
datasets \cite{SNeIa} containing $580$ points. The chi square formula for
the supernovae datasets is given by,

\begin{equation}  \label{chisn}
\chi _{SN}^{2}(\mu _{0},p_{s})=\sum\limits_{i=1}^{580}\frac{[\mu
_{th}(\mu_{0},p_{s},z_{i})-\mu _{obs}(z_{i})]^{2}}{\sigma _{\mu (z_{i})}^{2}}%
,
\end{equation}
where, $\mu _{th}$ and $\mu _{obs}$ are respectively, the theoretical and
observed distance modulus with the standard error in the observed value
denoted by $\sigma _{\mu (z_{i})}$. The distance modulus $\mu (z)$ is
defined by $\mu (z)=m-M=5LogD_{l}(z)+\mu _{0}, $where $m$ and $M$ are
respectively, the apparent and absolute magnitudes of a standard candle. The
luminosity distance $D_{l}(z)$ and the nuisance parameter $\mu _{0}$ are
defined by $D_{l}(z)=(1+z)H_{0}\int_{0}^{z}\frac{1}{H(z^{\ast })}dz^{\ast }$
and $\mu _{0}=5Log\Big(\frac{H_{0}^{-1}}{Mpc}\Big)+25$ respectively. In
order to calculate luminosity distance, we have restricted the series of $%
H(z)$ upto tenth term then integrate the approximate series to obtain the
luminosity distance.

\subsection{BAO datasets sample}

\qquad Baryon Acoustic Oscillation (BAO) measures the structures in the
universe from very large scales. For BAO, we have used the datasets of $%
\frac{d_{A}(z_{\ast })}{D_{V}(z_{BAO})}$ \cite{gio}, where $z_{\ast }$ is
the photon decoupling redshift (according to Planck 2018 results \cite%
{Hz-Plank}, $z_{\ast }=1091$), the comoving angular-diameter distance $%
d_{A}(z)=\int\limits_{0}^{z}\frac{dz^{^{\prime }}}{H(z^{^{\prime }})}$ and $%
D_{v}(z)=\left[ d_{A}(z)^{2}\frac{z}{H(z)}\right] ^{\frac{1}{3}}$. The
datasets used for our analysis consisteing of six points (from surveys of
SDSS(R) \cite{padn}, 6dF Galaxy survey \cite{6df}, BOSS CMASS \cite{boss}
and WiggleZ \cite{wig}) and given in the following Table I.

\begin{center}
\begin{tabular}{|c|c|c|c|c|c|c|}
\hline
$z_{BAO}$ & $0.106$ & $0.2$ & $0.35$ & $0.44$ & $0.6$ & $0.73$ \\ \hline
$\frac{d_{A}(z_{\ast })}{D_{V}(z_{BAO})}$ & $30.95\pm 1.46$ & $17.55\pm 0.60$
& $10.11\pm 0.37$ & $8.44\pm 0.67$ & $6.69\pm 0.33$ & $5.45\pm 0.31$ \\ 
\hline
\end{tabular}

Table I: Values of $\frac{d_{A}(z_{\ast })}{D_{V}(z_{BAO})}$ for different
of $z_{BAO}$.
\end{center}

The chi square for BAO datasets ($\chi _{BAO}^{2}$) is defined as,

\begin{equation}
\chi _{BAO}^{2}=X_{BAO}^{T}C_{BAO}^{-1}X_{BAO}  \label{bao-chi}
\end{equation}%
Where,%
\begin{equation*}
X_{BAO}=\left( 
\begin{array}{c}
\frac{d_{A}(z_{\star })}{D_{V}(0.106)}-30.95 \\ 
\frac{d_{A}(z_{\star })}{D_{V}(0.2)}-17.55 \\ 
\frac{d_{A}(z_{\star })}{D_{V}(0.35)}-10.11 \\ 
\frac{d_{A}(z_{\star })}{D_{V}(0.44)}-8.44 \\ 
\frac{d_{A}(z_{\star })}{D_{V}(0.6)}-6.69 \\ 
\frac{d_{A}(z_{\star })}{D_{V}(0.73)}-5.45%
\end{array}%
\right) \,,
\end{equation*}

and the inverse covariance matrix is given by, 
\begin{equation*}
C^{-1}=\left( 
\begin{array}{cccccc}
0.48435 & -0.101383 & -0.164945 & -0.0305703 & -0.097874 & -0.106738 \\ 
-0.101383 & 3.2882 & -2.45497 & -0.0787898 & -0.252254 & -0.2751 \\ 
-0.164945 & -2.454987 & 9.55916 & -0.128187 & -0.410404 & -0.447574 \\ 
-0.0305703 & -0.0787898 & -0.128187 & 2.78728 & -2.75632 & 1.16437 \\ 
-0.097874 & -0.252254 & -0.410404 & -2.75632 & 14.9245 & -7.32441 \\ 
-0.106738 & -0.2751 & -0.447574 & 1.16437 & -7.32441 & 14.5022%
\end{array}%
\right) \,.
\end{equation*}

\subsection{Results}

With the above samples, we have found the likelihood contours for our model
parameters $\alpha $ \& $\beta $ with $1$-$\sigma $, $2$-$\sigma $ and $3$-$%
\sigma $ errors in the $\alpha $-$\beta $ plane and shown in the following
figures. We have minimize the chi square for $OHD$ sample independently and
then combinedly as $OHD+BAO$, $OHD+SN$, $SN+BAO$ and finally $OHD+SN+BAO$
and summarized the constrained values in Table-2.

\begin{figure}[]
\subfloat[Contour plot for OHD\label{sfig:testa}]{
  \includegraphics[scale =0.49]{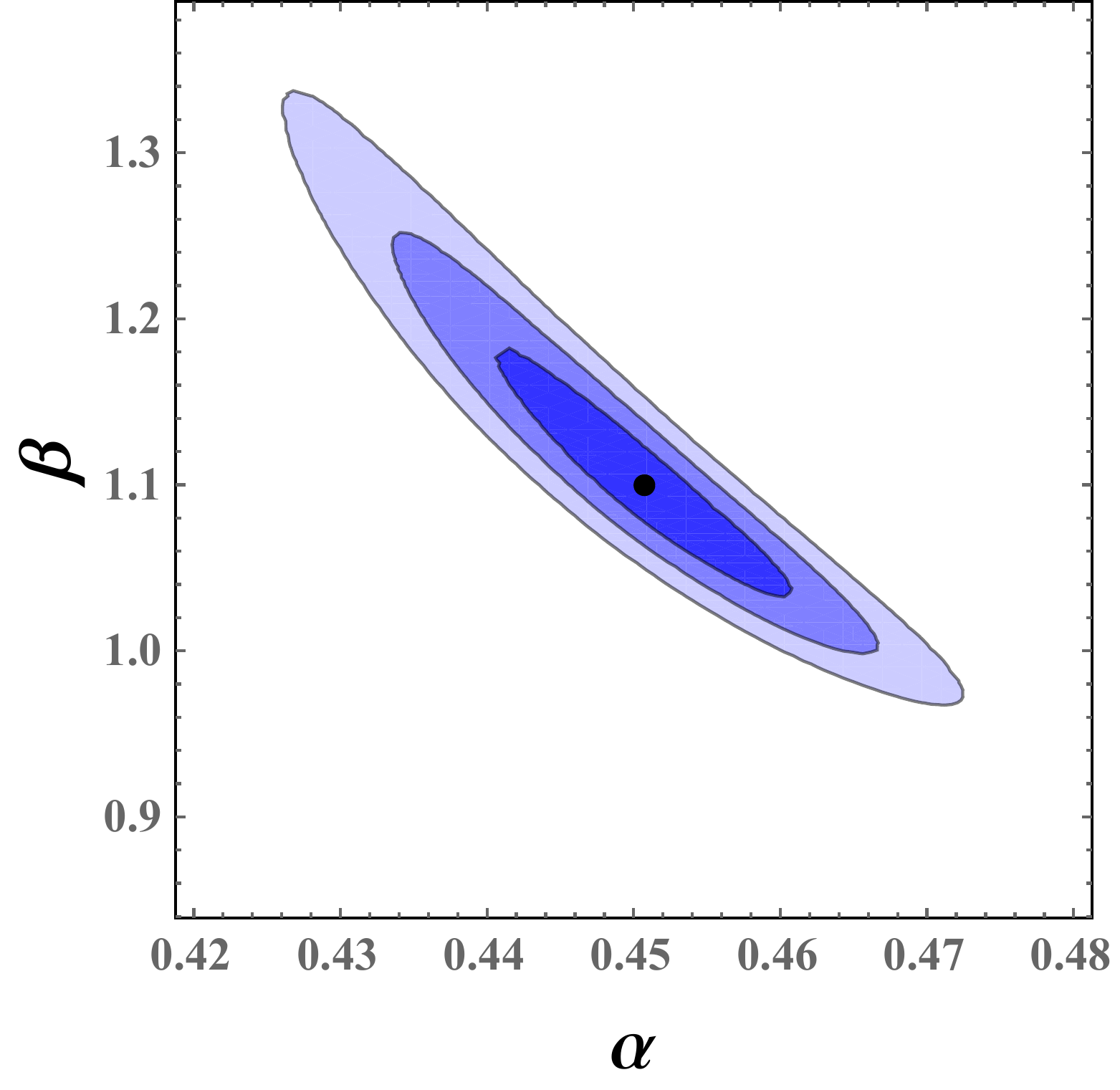}
}\hfill 
\subfloat[Contour plot for combined OHD+BAO\label{sfig:testa}]{
  \includegraphics[scale =0.59]{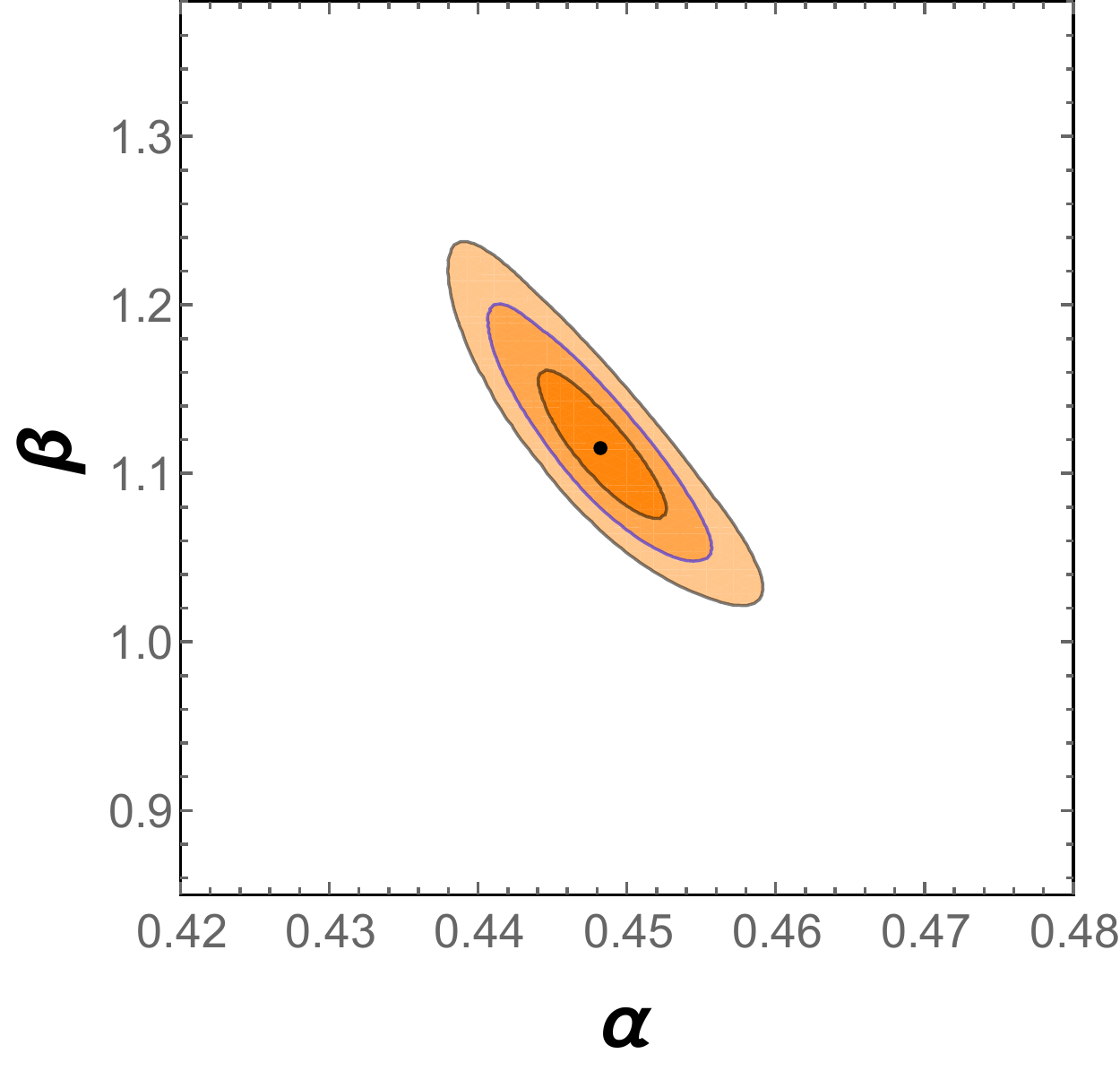}
}\hfill 
\subfloat[Contour plot for combined OHD+SN\label{sfig:testa}]{
  \includegraphics[scale =0.49]{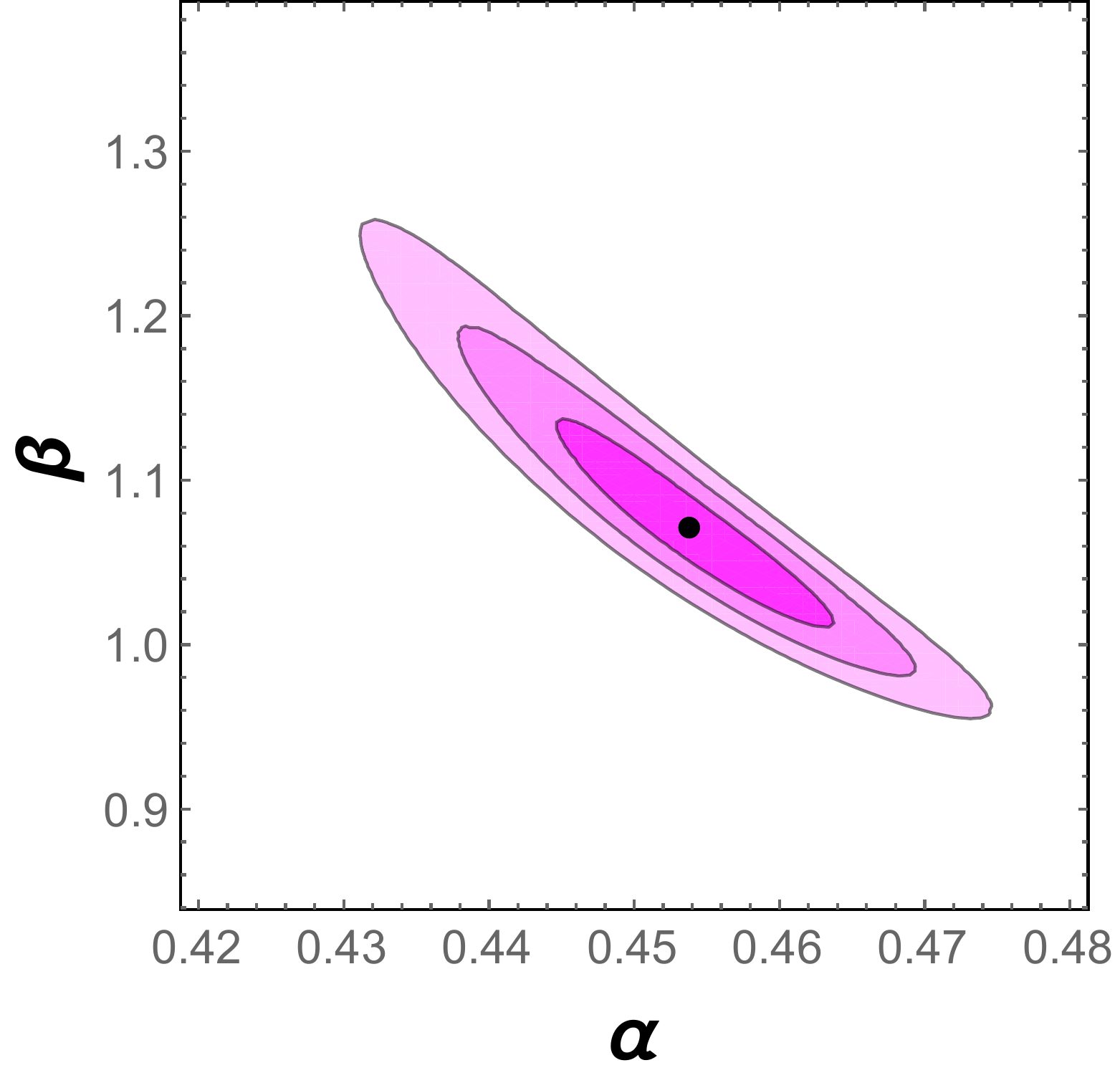}
}\hfill 
\subfloat[Contour plot for combined SN+BAO\label{sfig:testa}]{
  \includegraphics[scale =0.49]{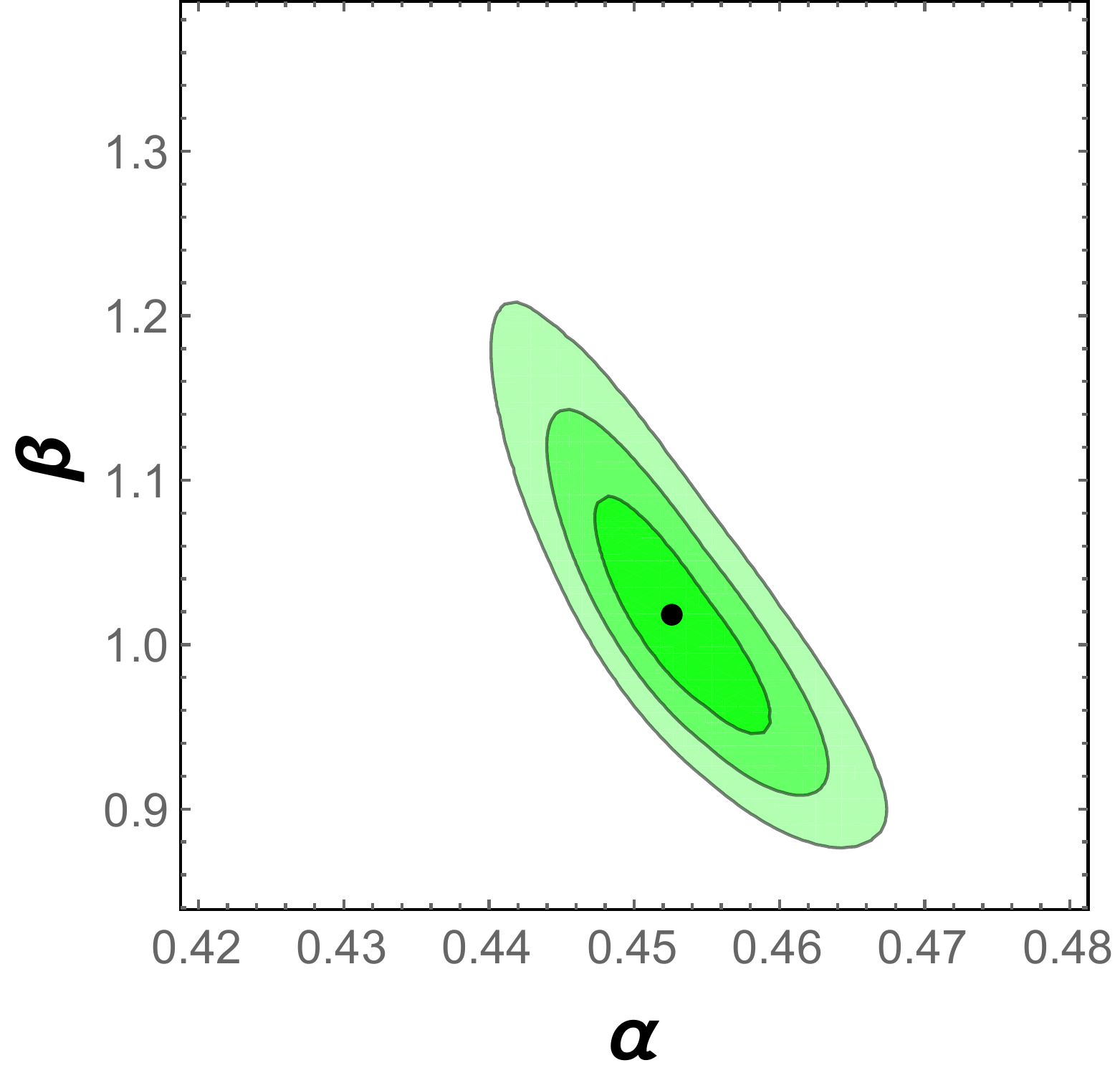}
}\hfill 
\subfloat[Contour plot for combined OHD+SN+BAO\label{sfig:testa}]{
  \includegraphics[scale =0.59]{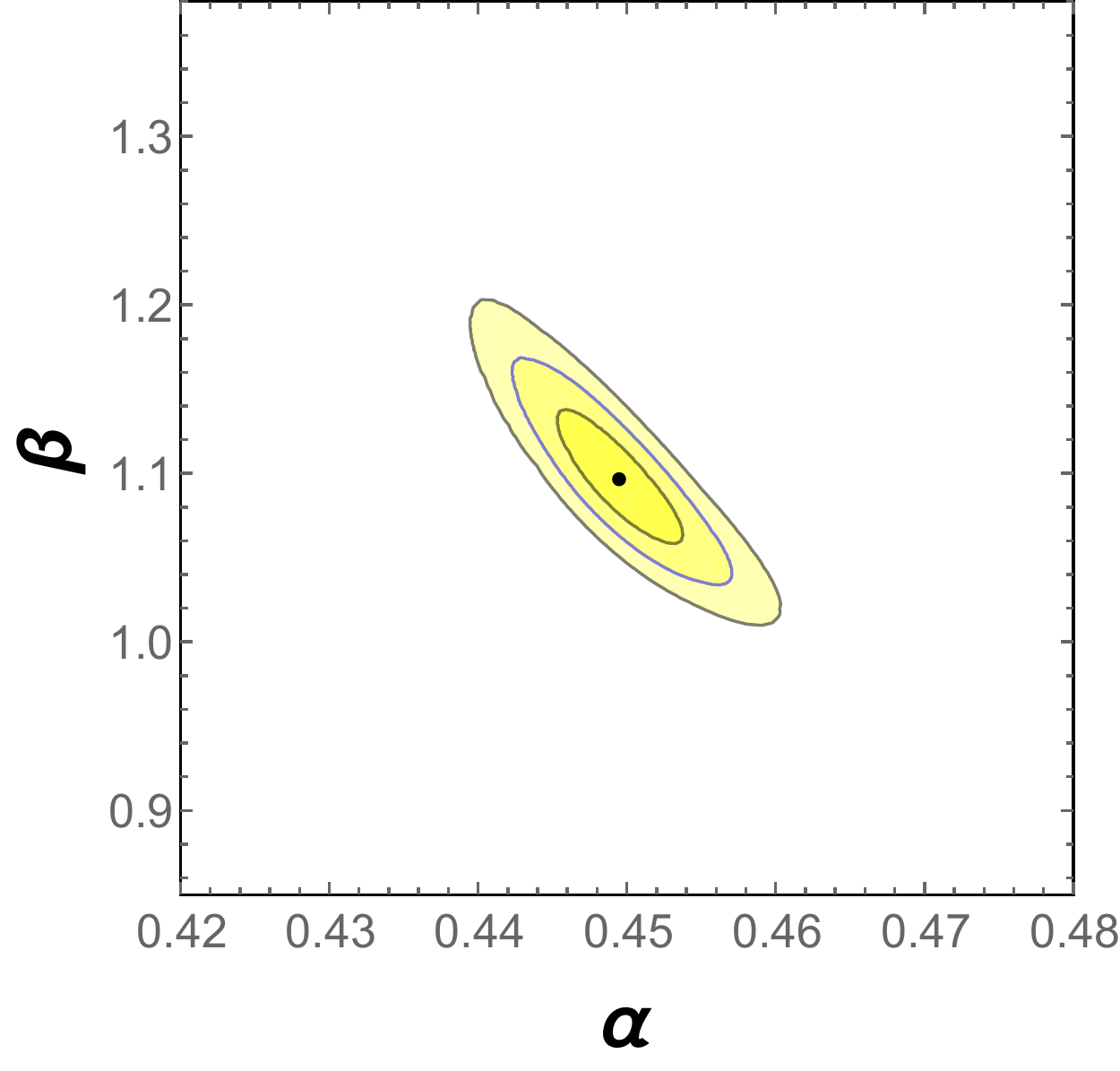}
}\hfill
\caption{Figures show the contour plots for likelihood values of the model
parameters $\protect\alpha$ \& $\protect\beta$ with samples of OHD, SN, BAO
and some combined datasets at 1-$\protect\sigma$, 2-$\protect\sigma$, 3-$%
\protect\sigma$ levels.}
\label{f1}
\end{figure}

We tabulate the constrained values of the model parameters $\alpha $ \& $%
\beta $ as follows together with the minimum chi square values. Also, we
calculate the values of deceleration parameter for different datasets at
present ($z=0$) which are obtained from Fig. \ref{f1}.

\begin{center}
Table.2: Constrained values of the model parameters with minimum chi square
values and the present values of deceleration parameter

\begin{tabular}{|l|c|c|c|c|c|}
\hline
$Datasets$ & $\alpha $ & $\beta $ & $\chi _{\min }^{2}/dof$ & $q(z=0)$ & $%
z_{t}$ \\ \hline
$H(z)$ & $0.45_{-0.02}^{+0.02}$ & $1.10_{-0.11}^{+0.11}$ & $0.574$ & $%
-0.50_{-0.11}^{+0.11}$ & $-0.76_{-0.37}^{+0.27}$ \\ \hline
$H(z)+BAO$ & $0.44_{-0.02}^{+0.02}$ & $1.11_{-0.11}^{+0.11}$ & $0.564$ & $%
-0.48_{-0.11}^{+0.13}$ & $-0.73_{-0.11}^{+0.13}$ \\ \hline
$H(z)+SN$ & $0.45_{-0.02}^{+0.02}$ & $1.07_{-0.11}^{+0.11}$ & $0.940$ & $%
-0.51_{-0.15}^{+0.23}$ & $-0.74_{-0.15}^{+0.23}$ \\ \hline
$SN+BAO$ & $0.45_{-0.01}^{+0.02}$ & $1.01_{-0.11}^{+0.11}$ & $0.966$ & $%
-0.57_{-0.16}^{+0.21}$ & $-0.85_{-0.60}^{+0.44}$ \\ \hline
$H(z)+SN+BAO$ & $0.44_{-0.01}^{+0.01}$ & $1.09_{-0.11}^{+0.11}$ & $0.936$ & $%
-0.50_{-0.11}^{+0.12}$ & $-0.67_{-0.36}^{+0.26}$ \\ \hline
\end{tabular}
\end{center}

\begin{figure}[]
\begin{center}
$%
\begin{array}{c@{\hspace{.1in}}c}
\includegraphics[width=3.0 in, height=2.5
in]{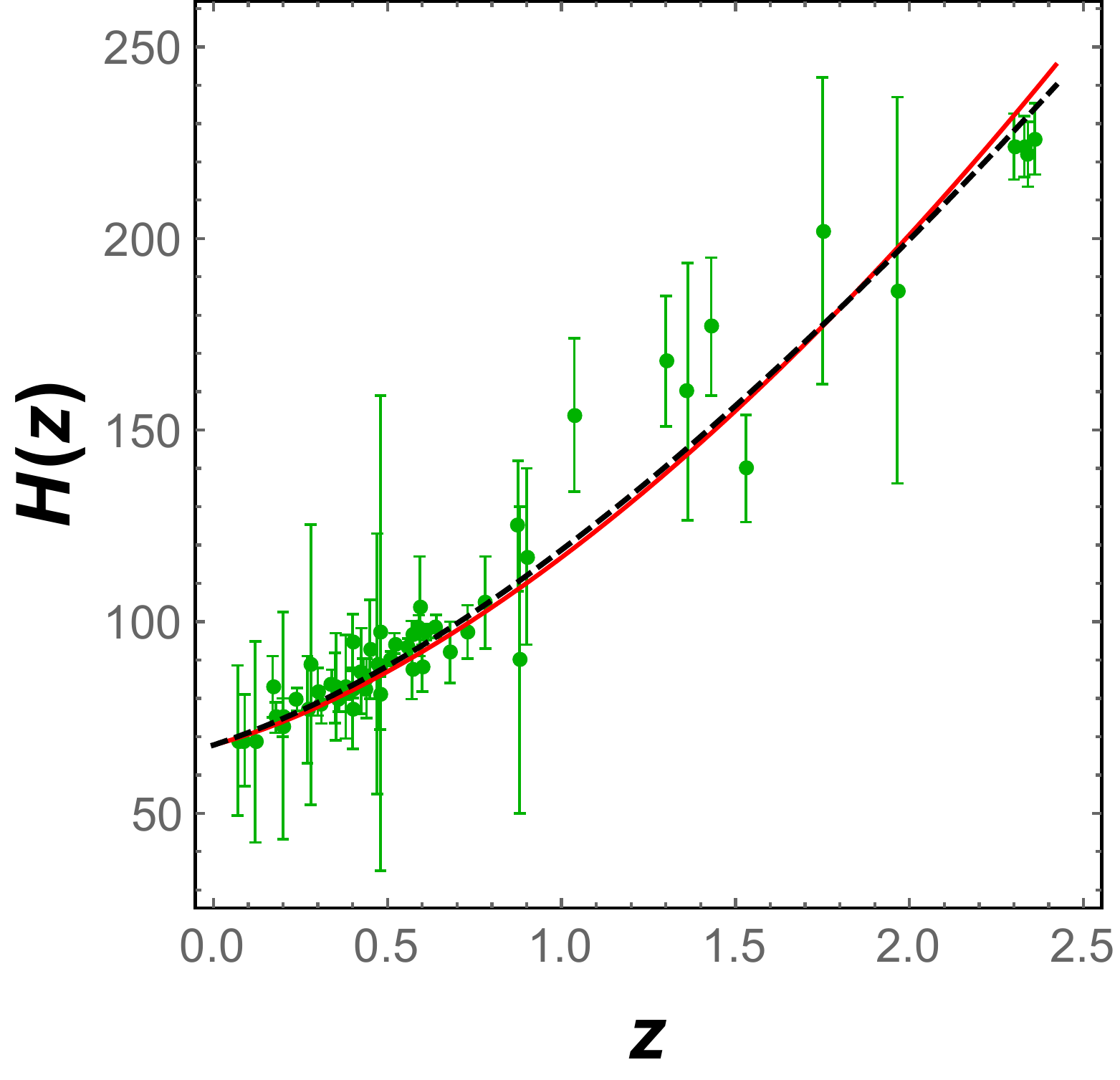} & 
\includegraphics[width=3.0 in,
height=2.5 in]{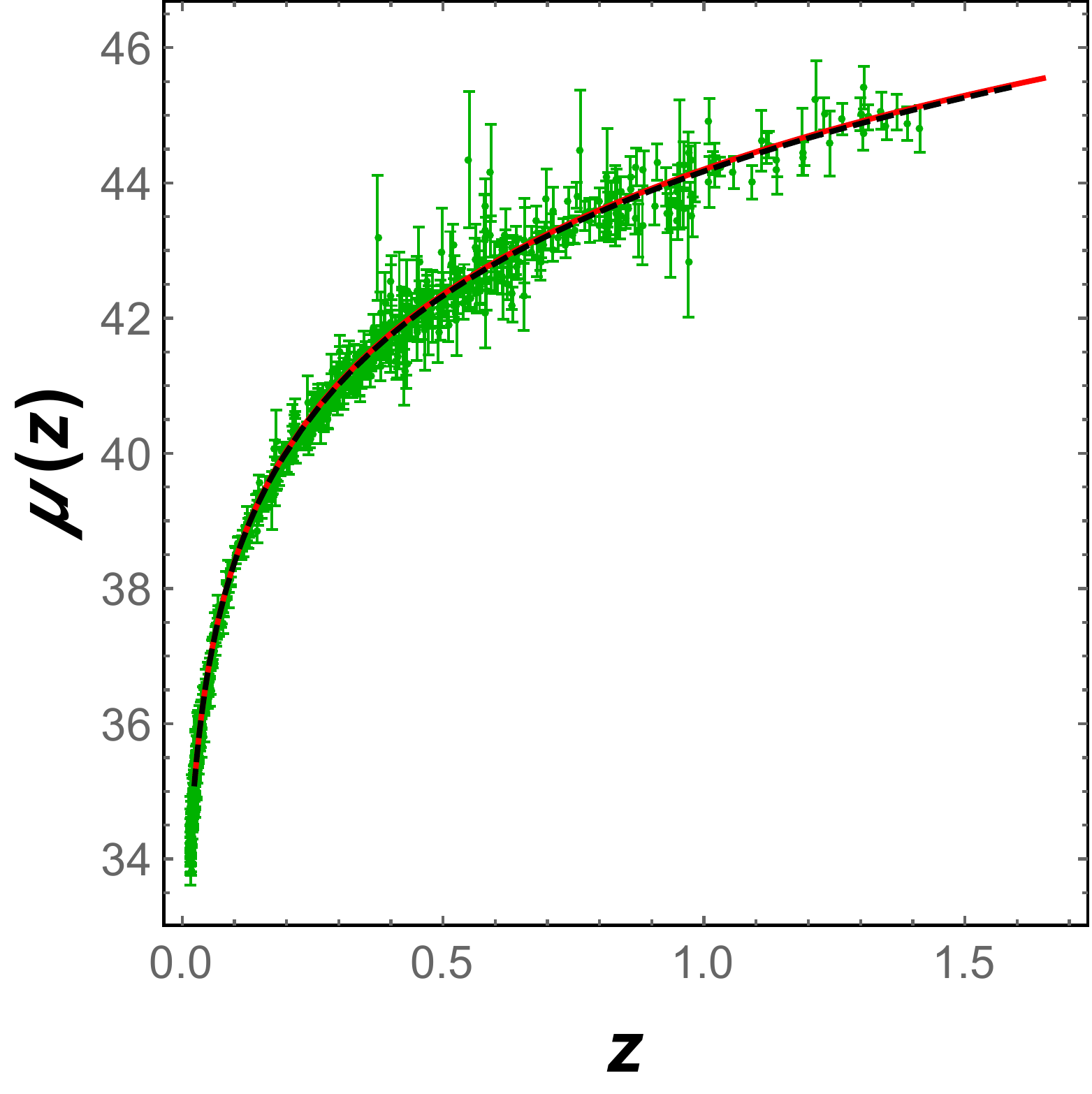} \\ 
\mbox (a) & \mbox (b)%
\end{array}
$%
\end{center}
\caption{Figures (a) and (b) are respectively the error bar plots for $57$
data points from OHD and for $580$ points from Union$2.1$ compilation
supernovae datasets. Solid red lines are presented model compared with $%
\Lambda $CDM model shown in black dashed lines in both the plots.}
\label{error}
\end{figure}

The error bar plots of the $OHD$ sample and the Union $2.1$ compilation
sample are shown in the Fig. \ref{error}.

\section{Evolution of cosmological and cosmographic parameters}\label{IV}

\begin{figure}[]
\centering
\includegraphics[width=8.5 cm]{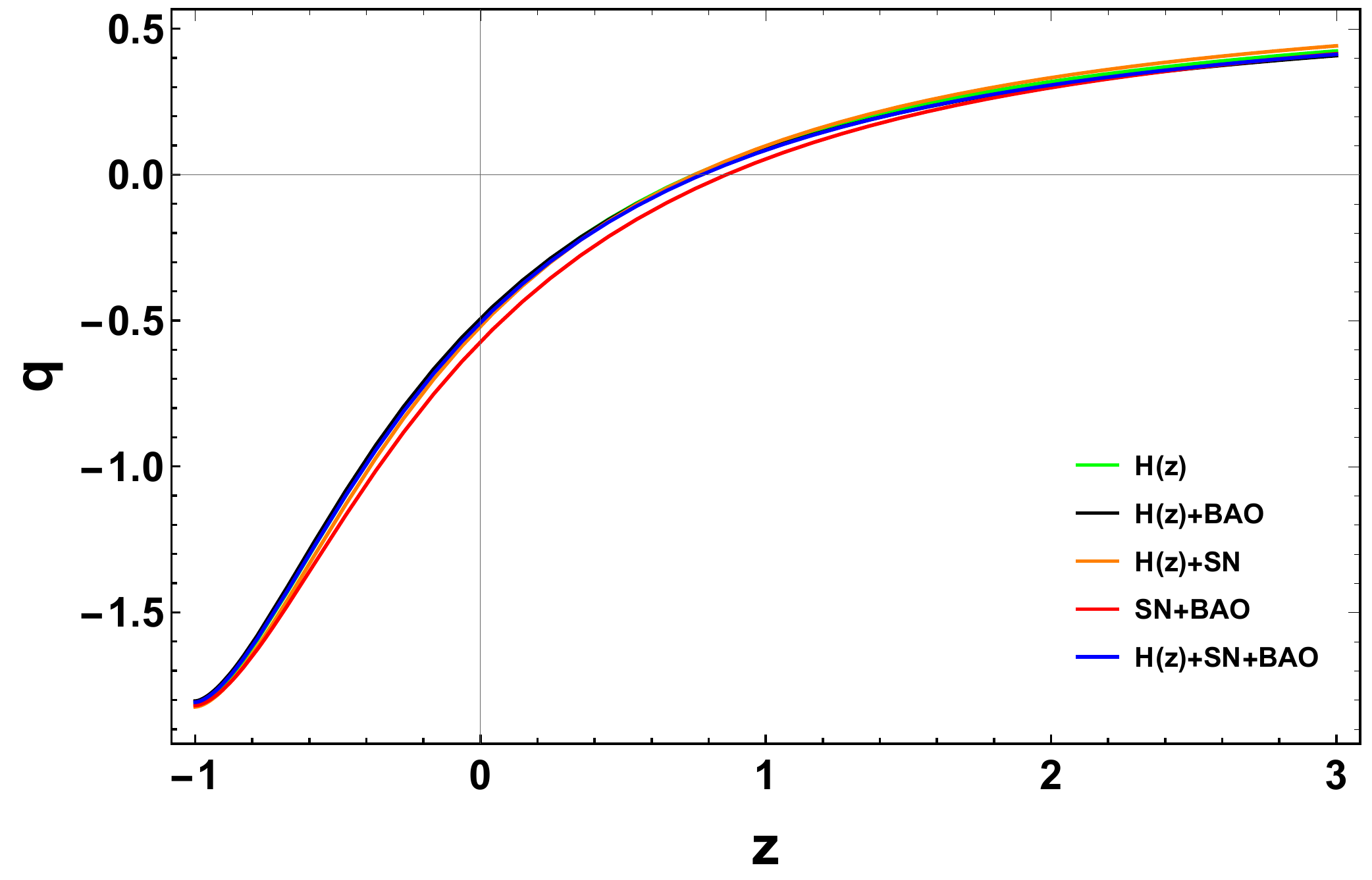}
\caption{Evolution of the deceleration parameter as a function of redshift
for different constrained values of model parameters $\protect\alpha$ \& $%
\protect\beta$ showing phase transition and present acceleration.}
\label{q(z)}
\end{figure}

\qquad The very useful way of describing an increasing or decreasing rate of
expansion of the universe is to study the deceleration parameter. The form
of deceleration parameter considered here (see equation (\ref{17})) contain
two parameters which are constrained through some datasets, so we can now
discuss it's evolution with the numerical values. The following plot shows
the evolution of $q$ w.r.t. redshift $z$ that explains it's evolution in the
near past, present evolution and the signature flipping behavior for the
above constrained values of the model parameters $\alpha $ and $\beta $ (see
Fig. \ref{q(z)}). As we know, although the negative value of q corresponds
to the accelerated period, the positive q refers to the decelerating phase.
We can see from the figure \ref{q(z)} that the deceleration parameter q
varies with z from positive to negative. This demonstrates a transition from
early deceleration to the universe's present acceleration.

If we talk about equation of state parameter (EoS), then we see that it
reflects the relation of energy density and pressure which basically
connects to the evolution of the universe. The EoS parameter for radiation
dominated phase is illustrated by $\omega =\frac{1}{3}$ followed by the dust
phase i.e matter dominated phase with $\omega =0$. The cosmological constant
is represented by $\omega =-1$ which is also known as $\Lambda $CDM model.
Also $-1<\omega <0$, shows the quintessence phase whereas $\omega <-1$ is
the phantom stage.

Now, using equations (\ref{16}) and (\ref{17}) in equations (\ref{10}), (\ref%
{11}) and (\ref{8}), we have the expressions for the pressure, the energy
densities and the potential function of the scalar field can be written as,

\begin{equation}
p_{eff}=-H_{0}^{2}\left( 2\frac{(\beta (z+1))^{8\alpha ^{2}}+\alpha
^{2}\left( 4-8(\beta (z+1))^{8\alpha ^{2}}\right) +1}{(\beta (z+1))^{8\alpha
^{2}}+1}+1\right) \frac{\left( (\beta (z+1))^{8\alpha ^{2}}+1\right) ^{3}}{%
\left( \beta ^{8\alpha ^{2}}+1\right) ^{3}(z+1)^{8\alpha ^{2}}}\text{.}
\label{peff}
\end{equation}

\begin{equation}
\rho _{eff}=3H_{0}^{2}\frac{\left( (\beta (z+1))^{8\alpha ^{2}}+1\right) ^{3}%
}{\left( \beta ^{8\alpha ^{2}}+1\right) ^{3}(z+1)^{8\alpha ^{2}}}\text{.}
\label{rhoeff}
\end{equation}

\begin{equation}
\rho _{\phi }=3H_{0}^{2}\frac{\left( (\beta (z+1))^{8\alpha ^{2}}+1\right)
^{3}}{\left( \beta ^{8\alpha ^{2}}+1\right) ^{3}(z+1)^{8\alpha ^{2}}}%
-3H_{0}^{2}\Omega _{m_{0}}(1+z)^{3}\text{. }  \label{rhophi}
\end{equation}%
where $\Omega _{m_{0}}=\frac{\rho _{0}}{3H_{0}^{2}}$. The expression for the
potential function $V(\phi )$ of the scalar field is obtained as,

\begin{equation}
V(\phi )=H_{0}^{2}\left( 2+\frac{(\beta (z+1))^{8\alpha ^{2}}+\alpha
^{2}\left( 4-8(\beta (z+1))^{8\alpha ^{2}}\right) +1}{(\beta (z+1))^{8\alpha
^{2}}+1}\right) \frac{\left( (\beta (z+1))^{8\alpha ^{2}}+1\right) ^{3}}{%
\left( \beta ^{8\alpha ^{2}}+1\right) ^{3}(z+1)^{8\alpha ^{2}}}%
-1.5H_{0}^{2}\Omega _{m_{0}}(1+z)^{3}\text{.}  \label{poten}
\end{equation}

The evolution of the effective energy density of the matter energy density,
scalar field energy density and the scalar field potential are shown in the
figure.

\begin{figure}[tbp]
\begin{center}
$%
\begin{array}{c@{\hspace{.1in}}c}
\includegraphics[width=3.0 in, height=2.5 in]{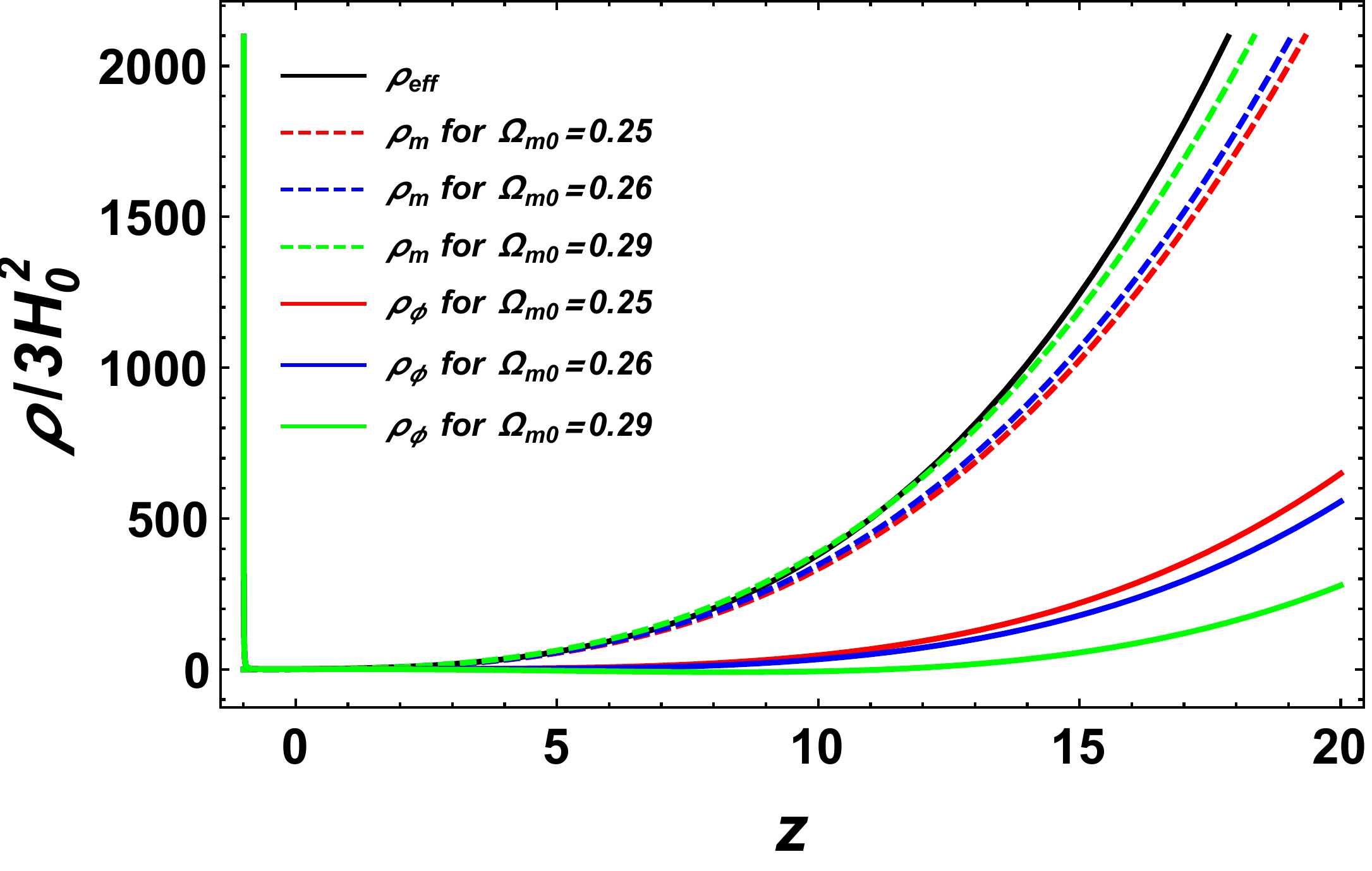} & %
\includegraphics[width=3.0 in, height=2.5 in]{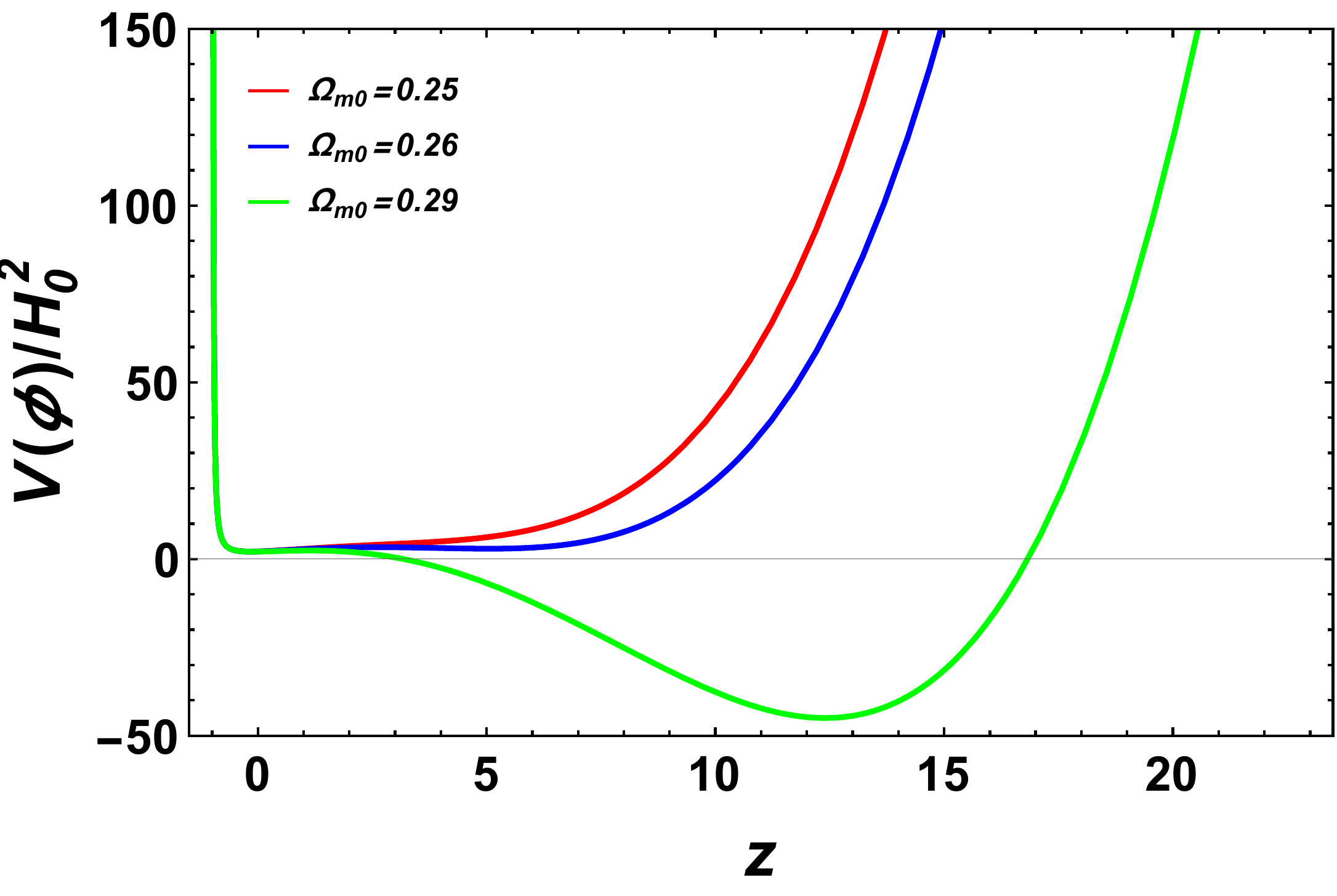} \\ 
\mbox (a) & \mbox (b)%
\end{array}
$%
\end{center}
\caption{In this figure panel (a) shows the evolution of effective energy
density, matter energy density and scalar field energy density as a function
of redshift while panel (b) shows the evolution of scalar field potential $V(%
\protect\phi )$ as function of redshift.}
\label{N-vphi.pdf}
\end{figure}

\begin{figure}[tbp]
\begin{center}
$%
\begin{array}{c@{\hspace{.1in}}c}
\includegraphics[width=3.0 in, height=2.5 in]{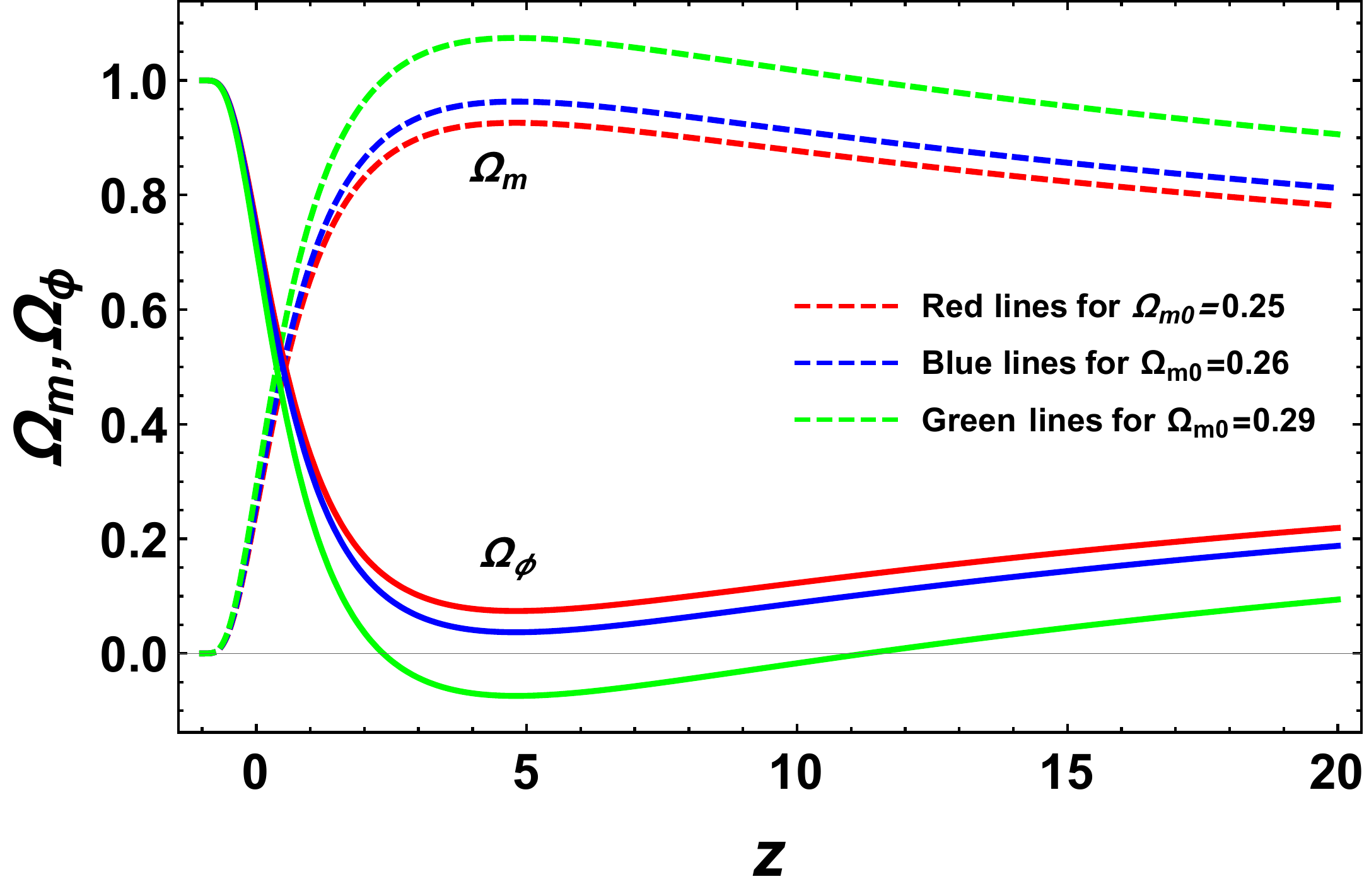} & 
\end{array}
$%
\end{center}
\caption{The figure shows the evolution of the density parameters for matter
($\Omega _{m}$) and the scalar field ($\Omega _{\protect\phi }$) as
functions of redshift $z$.}
\label{Omega m phi-2021i.pdf}
\end{figure}

Now, the effective equation of state parameter $\omega _{eff}=\frac{p_{eff}}{%
\rho _{eff}}$ reads as,

\begin{equation}
\omega _{eff}=-\frac{1}{3}\left( 2\frac{(\beta (z+1))^{8\alpha ^{2}}+\alpha
^{2}\left( 4-8(\beta (z+1))^{8\alpha ^{2}}\right) +1}{(\beta (z+1))^{8\alpha
^{2}}+1}+1\right) .  \label{weff}
\end{equation}

Moreover, with the dark energy domination, the expression for the equation
of state of scalar field or the dark energy $\omega _{\phi }$ can be written
as,

\begin{equation}
\omega _{\phi }=\frac{1}{3}\frac{\frac{\left( (\beta (1+z))^{8\alpha
^{2}}+1\right) ^{3}}{\left( \beta ^{8\alpha ^{2}}+1\right)
^{3}(1+z)^{8\alpha ^{2}}}\left( 2\frac{(\beta (z+1))^{8\alpha ^{2}}+\alpha
^{2}\left( 4-8(\beta (z+1))^{8\alpha ^{2}}\right) +1}{(\beta (z+1))^{8\alpha
^{2}}+1}+1\right) }{\Omega _{m0}(1+z)^{3}-\frac{\left( (\beta
(1+z))^{8\alpha ^{2}}+1\right) ^{3}}{\left( \beta ^{8\alpha ^{2}}+1\right)
^{3}(1+z)^{8\alpha ^{2}}}}.  \label{w}
\end{equation}

\begin{figure}[]
\begin{center}
$%
\begin{array}{c@{\hspace{.1in}}c}
\includegraphics[width=3.0 in, height=2.5 in]{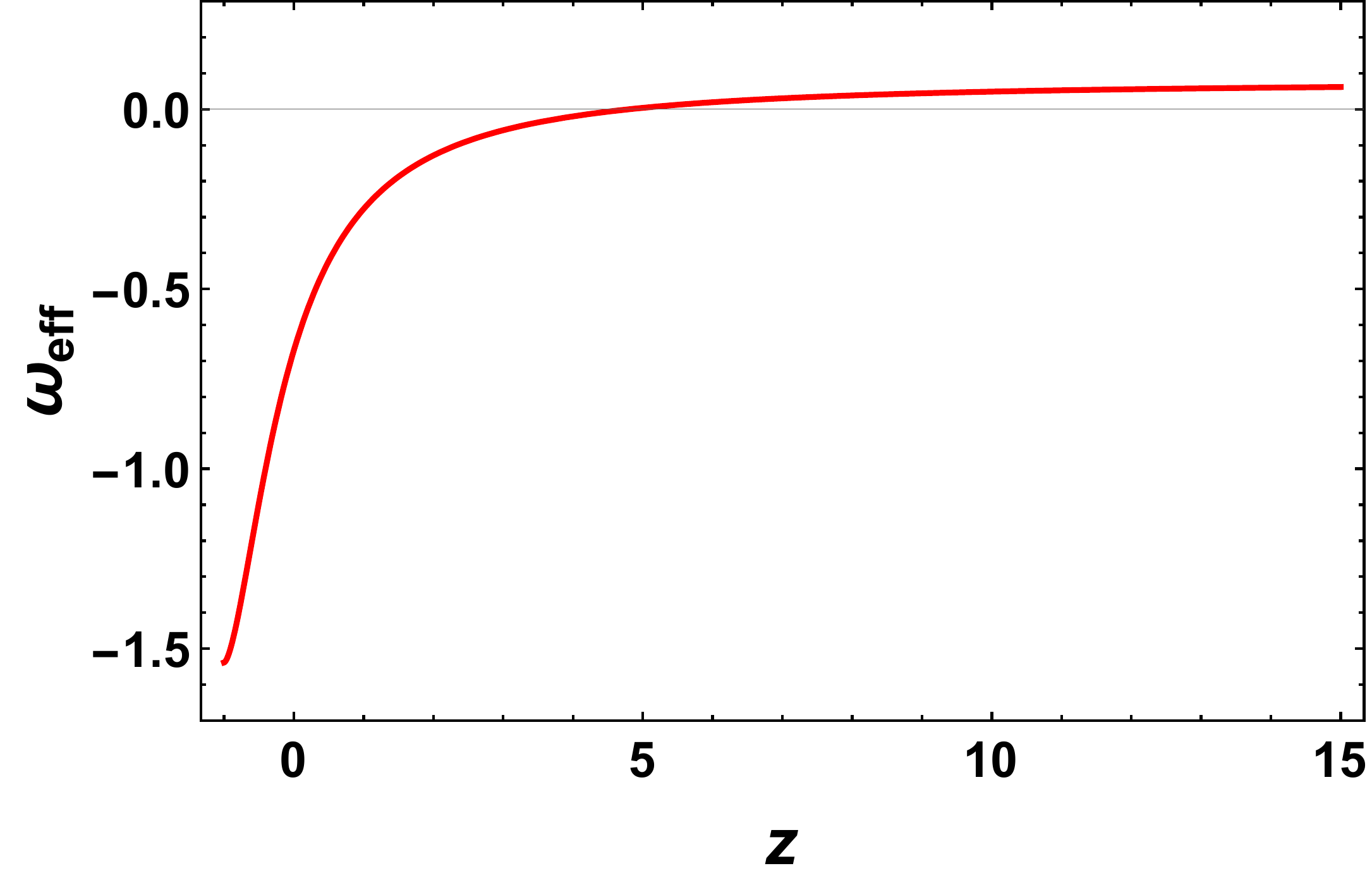} & %
\includegraphics[width=3.0 in, height=2.5 in]{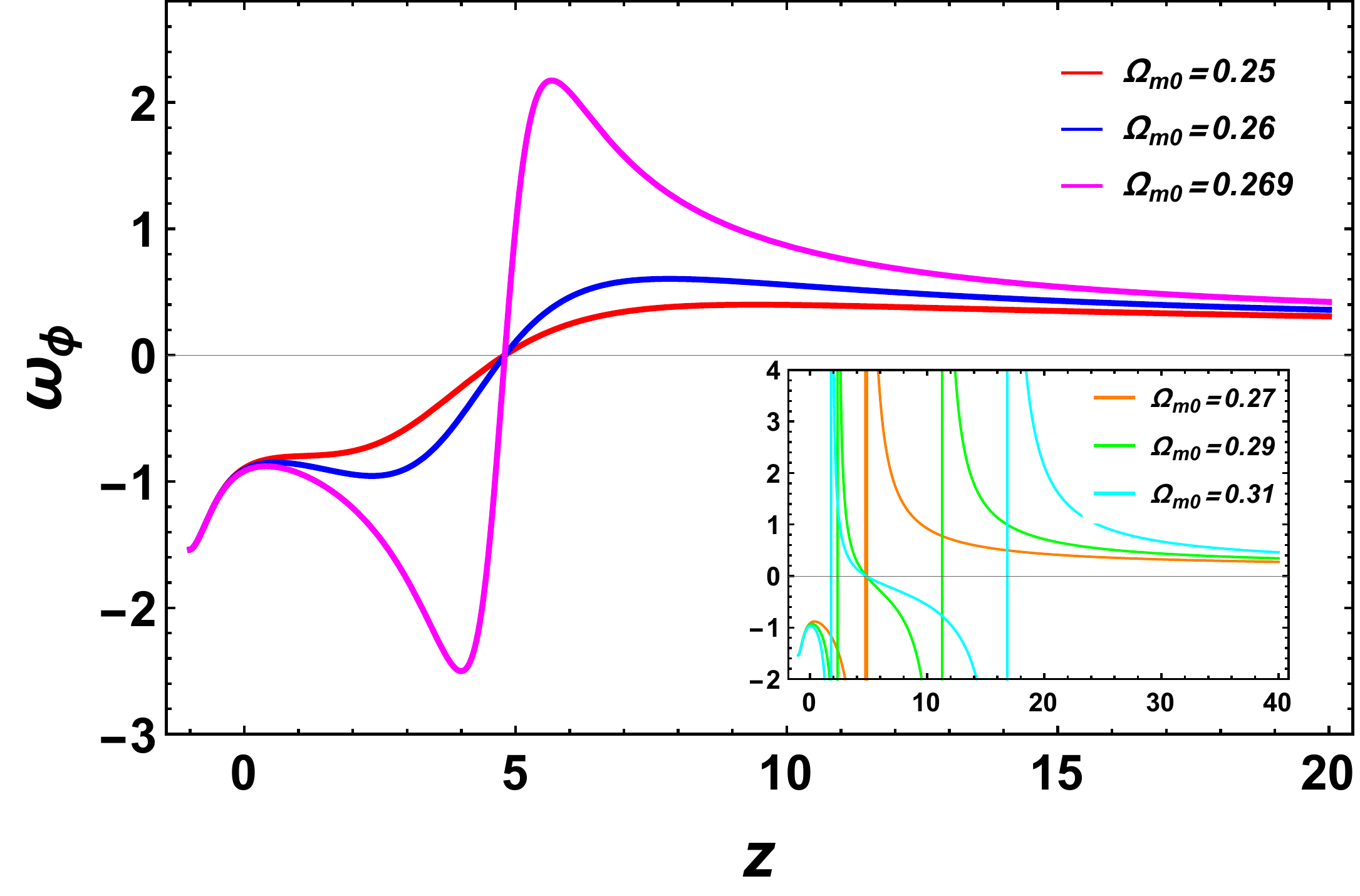} \\ 
\mbox (a) & \mbox (b)%
\end{array}
$%
\end{center}
\caption{In this figure, the panel (a) shows the evolution of effective EoS
parameter ($\protect\omega ^{eff}$) for the constrained values of $\protect%
\alpha=0.449451$ and $\protect\beta =1.096408$ and the panel (b) shows the
evolution of scalar field (or dark energy) EoS parameter ($\protect\omega _{%
\protect\phi } $) as functions of redshift.}
\label{N-wphi.pdf}
\end{figure}

In Fig \ref{N-vphi.pdf}, the energy density $\rho_{eff}$, $\rho_{m}$ and $%
\rho_{\phi}$ is showing positive behavior for according to different value
of $\Omega_{m_{0}}$. It is seen that the energy density is higher for the
effective energy density of the matter field if compared with energy density
in the presence of scalar field. Also the evolution of scalar field
potential is shown which is responsible for negative pressure.\newline

The evolution density parameter for matter and scalar field is shown in Fig%
\ref{Omega m phi-2021i.pdf} for different values of $\Omega_{m_{0}}$. Also,
in Figs \ref{N-wphi.pdf}, the behavior of $\omega _{eff}$ and $\omega _{\phi}
$ are shown for the constrained values of $\alpha=0.449451$ and $\beta
=1.096408$ and some chosen values values of the matter density parameter $%
\Omega _{m_{0}}$ as shown in the figure. This analysis also put constrain on
the $\Omega _{m_{0}}$ value which should be $\leq 0.269$ for smooth
evolution of the scalar field (or dark energy) EoS parameter. The curve
shows a negative behavior at early times and evolve through different phases
of acceleration, deceleration and finally to phantom phase. It can also be
seen that for some values of $\Omega _{m_{0}}$, the curves shows a
singularity. The transition from one phase to the other indicates the
possibility for evolution of the universe. Planck observations are known to
be considered the best approximations of cosmological effects. Planck's 2018
findings indicate that the Hubble constant is $H_{0}=67.8\pm 0.9$ km $s^{-1}
Mpc^{-1}$ and $\Omega_{m}=0.308\pm 0.012$, but the obtained value of $%
\Omega_{m_{0}}$ is 0.269. This deviation point towards a tension with Planck
results. \newline

Moreover, the higher order derivatives of deceleration parameter $q$ such as
jerk ($j$), snap ($s$) and lerk ($l$) are important in understanding the
past and future evolution of the universe \cite{sanjay}. They are
represented as \cite{Pan}: 
\begin{equation}  \label{eqj}
j(z)=(1+z)\frac{dq}{dz}+q(1+2q),
\end{equation}

\begin{eqnarray}  \label{eqs}
s(z)= -(1+z)\frac{dj}{dz}-j(2+3q),
\end{eqnarray}

\begin{eqnarray}  \label{eql}
l(z)=-(1+z)\frac{ds}{dz}-s(3+4q).
\end{eqnarray}

\begin{figure}[]
\centering
\includegraphics[width=8.5 cm]{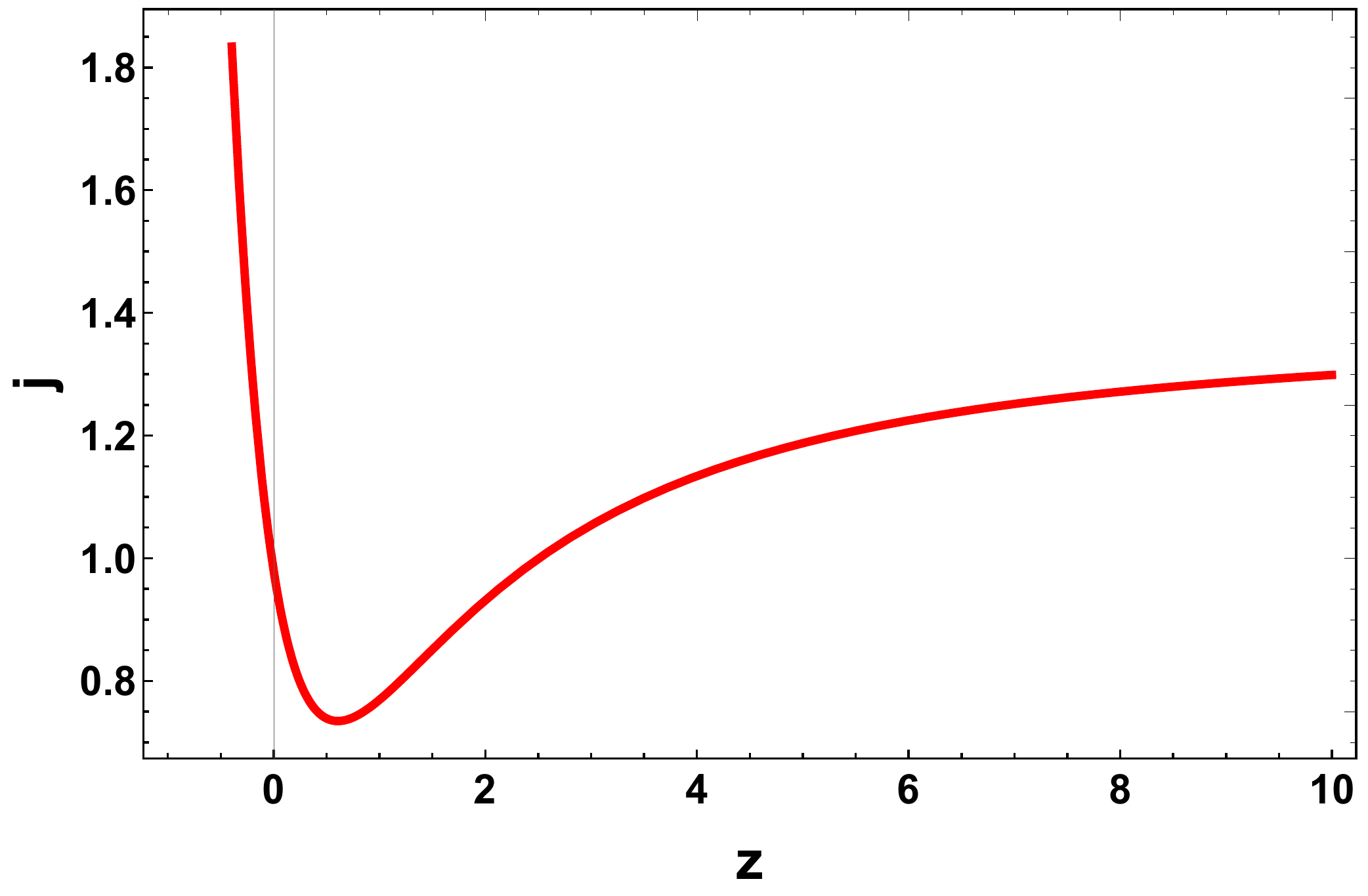}
\caption{{\protect\scriptsize The plots of jerk $j$, snap $s$ and lerk $l$
parameters \textit{\ vs.} redshift $z$ for the constrained values of $%
\protect\alpha=0.449451$ and $\protect\beta =1.096408$.}}
\label{jerk}
\end{figure}

These higher order derivatives can be useful in understanding the future
evolution of the universe owing to the fact that $q(z)$ can be strictly
constrained from observations. The jerk parameter $j$ is related to the
third time derivative of $a$ as a higher-order derivative of the scale
component. The higher-order derivatives can describe the dynamics of the
universe and may be connected to the appearance of abrupt future
singularities \cite{Pan,Dabrowski}. In the statefinder diagnostic, the jerk
parameter $j$ is often used to discriminate against various dark energy or
modified gravity models. Zhai \cite{Zhai} proposed different kinds of
parameterizations of $j$ as a function of the redshift z. A vital feature of 
$j$ is that for the $\Lambda$CDM model, $j = 1$ always. The deviation from $%
j=1$ enables us to constrain the departure from the $\Lambda$CDM value.. The
value of $j_{0}$ according to constrain value of $\alpha$ and $\beta$ is $%
j_{0}=-0.98_{-0.02}^{+0.06}$ \cite{Mamon}. The behavior of $j(z)$ is shown
in Figs. \ref{jerk}. It can be clearly observed that $j(z)$ increases with
redshift indicating a decelerated phase in the past and an accelerated phase
in future. Interestingly the kinematic quantity $j(z)$ is positive which is
a reminiscent of an accelerated expansion. Also note that $j(z)\neq 1$ at $%
z=0$ which does not correspond to $\Lambda$CDM cosmology. As mentioned in 
\cite{sanjay}, this can be thought as an expansion caused purely due to
modifications of gravity.

\section{Statefinder diagnostics}

\label{V}

\qquad The statefinder pairs $\{r,s\}$ and $\{r,q\}$ are the geometrical
quantities formulated directly from the metric and are employed to identify
various dark energy model. In the literature, the $\{r,s\}$ and $\{r,q\}$
pairs are defined as \cite{Statefinder1}. 
\begin{equation}  \label{d1}
q=-\frac{\ddot{a}}{aH^{2}}\text{, \ \ }r=\frac{\dddot{a}}{aH^{3}}\text{, \ \ 
}s=\frac{r-1}{3(q-\frac{1}{2})}\text{.}
\end{equation}%
The statefinder diagnostic is an useful tool in modern day cosmology and
being used to serve the purpose of distinguishing different dark energy
models \cite{Statefinder2}. In this setup, different trajectories in $r-s$
and $r-q$ planes define the temporal evolution for various dark energy
models. In a spatially flat FLRW background, the statefinder pair are
respectively $\{r,s\}=\{1,0\}$ and $\{1,1\}$ for $\Lambda $CDM and standard
cold dark matter (SCDM). In the $r-s$ and $r-q$ planes, the departure of any
dark energy model from these fixed points are analyzed. The pairs $\{r,s\}$
and $\{r,q\}$ for our model are shown in figure below.

\begin{figure}[]
\begin{center}
$%
\begin{array}{c@{\hspace{.1in}}c}
\includegraphics[width=3.0 in, height=2.5 in]{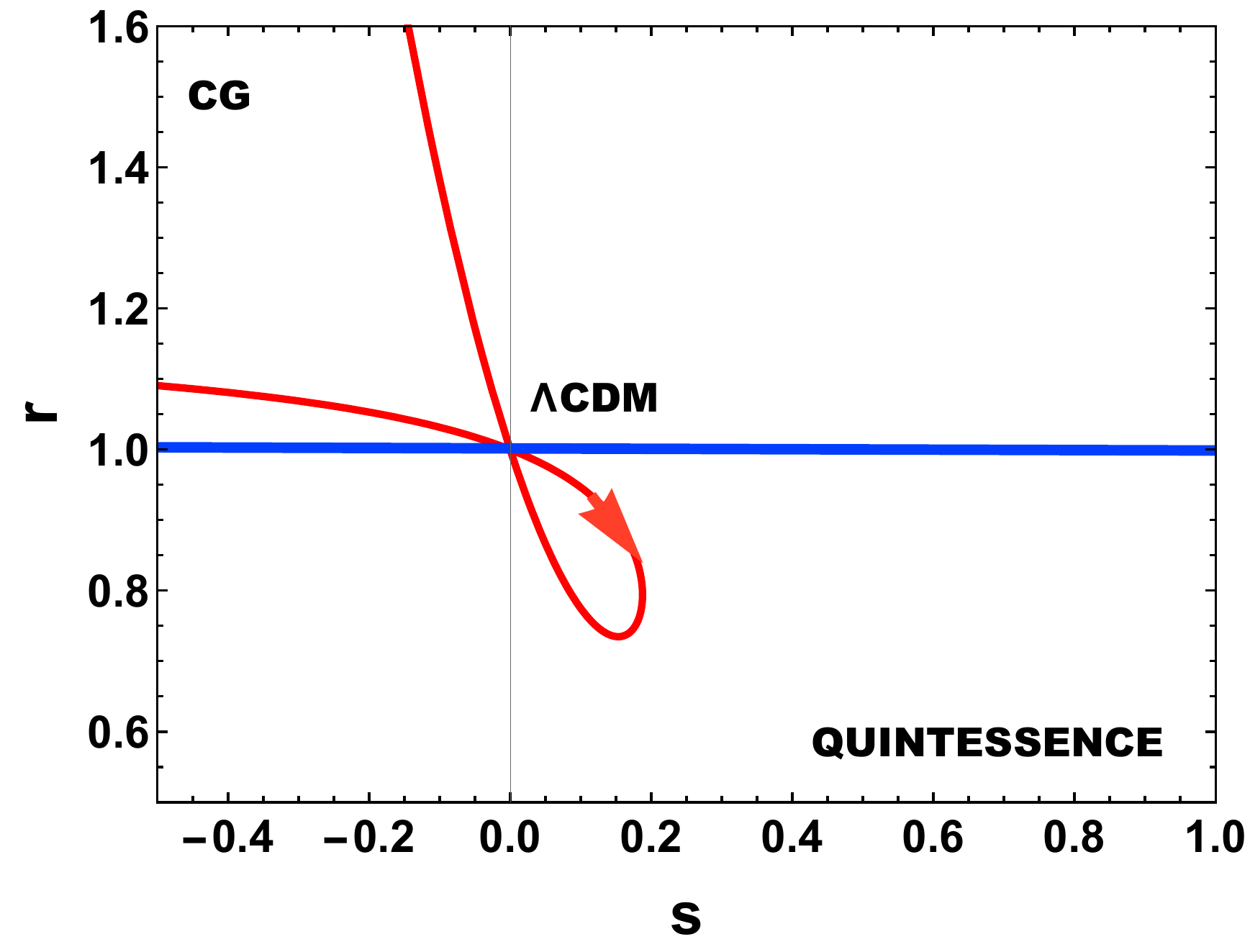} & %
\includegraphics[width=3.0 in, height=2.5 in]{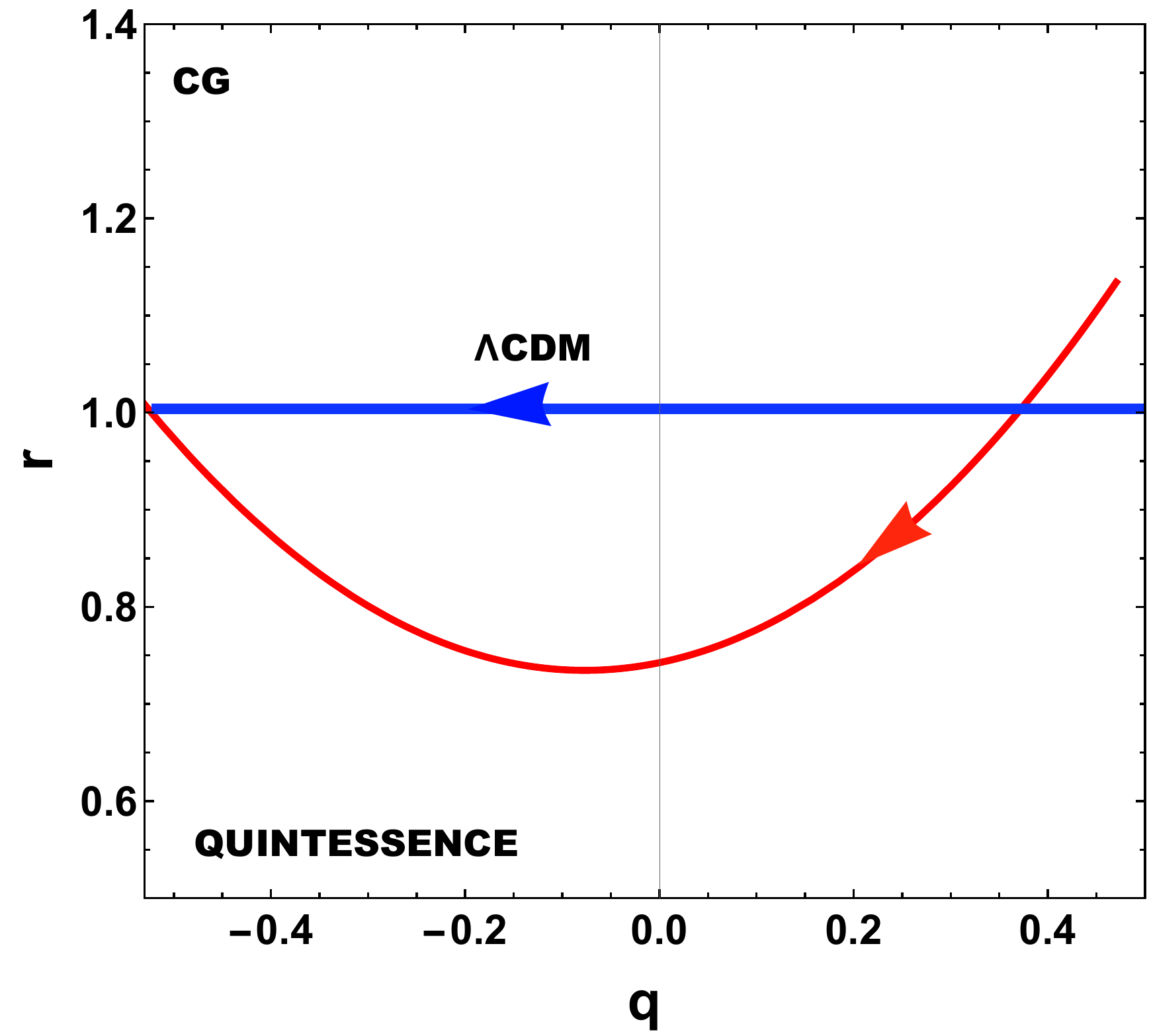} \\ 
\mbox (a) & \mbox (b)%
\end{array}
$%
\end{center}
\caption{Figures (a) and (b) shows the behavior of {r-s} and {r-q} plane in
red color respectively for the constrained values of $\protect\alpha%
=0.449451 $ and $\protect\beta =1.096408$.}
\label{r-q}
\end{figure}

In Fig. \ref{r-q}(a) we show the temporal evolution of the dark energy model
mimicked by our model. It is observed that at early times, the model
presumes values in the range $r>1$ and $s<0$ and therefore represents a
Chaplygin gas type dark energy model. Nonetheless, the model evolves into a
Quintessence type dark energy model at some point but then quickly reverts
back into CG gas at late times. Interestingly, it can be clearly observed
that throughout its temporal evolution, the model deviates significantly
from the point $\{r,s\} = \{1,0\}$. In Fig. \ref{r-q}(b) we show the
temporal evolution of our model in the $\{r,q\}$ to get additional
information regarding the parametrization. In this diagnostic plane, the
solid line in the middle depicts the evolution of the standard $\Lambda$CDM
cosmological model and also divides the plane into two equal halves with the
lower half belonging to Quintessence dark energy models and the upper half
to Chaplygin gas dark energy models. We clearly see that the profile starts
from the region $q>0$ and $r>1$ which corresponds to the SCDM universe. This
is then followed by the region $r<1$ and $q<0$ and finally approaches to
towards the de-Sitter phase with $r=1, q=-1$.

\section{Conclusion}

\label{VI}

In this paper we have considered the scalar field with positive potential to
study the accelerated expansion of the universe. To explain the late-time
acceleration, the EoS parameter $\omega$ must be negative as this implies
the cosmological pressure is negative. Negative cosmological pressure is the
hallmark of the presence of dark energy as this is only cosmic entity which
posses an anti-gravity effect. The potential $V(\phi)$ is responsible for
negative pressure. \newline

In order to solve the field equations, we took the assistance of a
supplementary equation since the value of $\rho _{\phi }$ and $p_{\phi }$ is
difficult to obtain. In this work, we employ a second degree parametrization
of deceleration parameter first proposed in \cite{Bakry and Shafeek}. We
used the latest 57 points of $H(z)$ dataset in the redshift range $0.07\leq
z\leq 2.42$, 580 points of SN data and BAO datasets to constrained the model
parameters using $\chi ^{2}$ minimization technique. The values of the
parameters $\alpha \ \&\ \beta $ considered here from all the datasets given
in Table 2, the present model shows a smooth transition for the deceleration
parameter (see Fig. \ref{q(z)}) from the deceleration ($q>0$) phase to the
present acceleration ($q<0$) phase of the Universe. It has been found that
the values of transition redshift $z_{t}$ (from decelerated phase to
accelerated expansion) and the values of present deceleration parameter $%
q_{0}$ depends on the model parameters $\alpha \ \&\ \beta $. It is
interesting to note that the values of $z_{t}=0.67_{-0.36}^{+0.26}$ and $%
q_{0}=-0.50_{-0.11}^{+0.12}$ obtained in our model are in good agreement
with the recent results as reported in \cite{Jesus/2020} and the references
therein. Further, we have studied the other kinematical parameter like jerk
parameter using the combined datasets ($H(z)+SN+BAO$). An alternative to
describing cosmological models similar to the $\Lambda $CDM model of
concordance is the jerk parameter. For LCDM model , the value of j is 1. The
deviation from $j=1$ enables us to constrain the departure from the $\Lambda 
$CDM value. According to the restricted value of $\alpha $ and $\beta $, the
value of $j_{0}=-0.98_{-0.02}^{+0.06}$ (close to 1)\textbf{.} The jerk
parameter increasing with respect to redshift indicating the accelerated
phase in the future Universe. \newline
Also, the evolution of $\omega _{eff}$ and $\omega _{\phi }$ is shown for
the constrained $\alpha =0.449451$ and $\beta =1.096408$ values and some of
the chosen $\Omega _{m_{0}}$ parameter density values. The transition from
positive to negative values indicates the decelerated to accelerated phase
of the universe. This study also constrains the value of $\Omega _{m_{0}}$
which should be $\leq 0.269$ for the smooth evolution of the EoS parameter
scalar field. It can also be shown that the curves display a singularity for
specific values of $\Omega _{m_{0}}$. It is understood that Planck
observations are regarded as the best approximations of cosmological
results. Planck's 2018 results suggest that $H_{0}=67.8\pm 0.9$ km $%
s^{-1}Mpc^{-1}$ and $\Omega _{m}=0.308\pm 0.012$ are the Hubble constants,
but the obtained value of $\Omega _{m_{0}}$ is 0.269 which results in
tension with Planck approximations.

Furthermore, to understand the parametrization from a cosmological point of
view, we also diagnose it geometrically using $\{r,s\}$, $\{r,q\}$ planes
and $Om(z)$ parameter. We observe that at early times, the model represents
a Chaplygin gas type dark energy model and later evolves into a Quintessence
type dark energy model at some point but then quickly reverts back into CG
gas at late times. Interestingly, the model deviates significantly from the
point $\{r,s\} = \{1,0\}$ and therefore do not coincide with $\Lambda$CDM
cosmology throughout the cosmic aeon.

\acknowledgements S. A. acknowledges CSIR, Govt. of India, New Delhi, for
awarding Junior Research Fellowship. PKS acknowledges CSIR, New Delhi, India
for financial support to carry out the Research project
[No.03(1454)/19/EMR-II Dt.02/08/2019]. We are very much grateful to the
honorable referee and to the editor for the illuminating suggestions that
have significantly improved our work in terms of research quality, and
presentation.

\end{document}